\documentclass[draft]{agujournal2019}
\usepackage{url} 
\usepackage{soul}
\usepackage[pdftex]{hyperref}
\usepackage{rotfloat}
\usepackage{amsmath,amssymb}
\usepackage{graphicx}
\usepackage{lscape}
\usepackage{float}
\usepackage{natbib}
\providecommand{\keywords}[1]{\textbf{\textit{Keywords: }} #1}

\draftfalse

\journalname{Astrobiology}

\begin{document}

%
%

\title{UV Transmission in Natural Waters on Prebiotic Earth}

%
%




\authors{S. Ranjan\affil{1,2,3,*}, C. L. Kufner\affil{4}, G. G. Lozano\affil{4}, Z. R. Todd\affil{4,5}, A. Haseki\affil{1,6}, D. D. Sasselov\affil{4}}


\affiliation{1}{Massachusetts Institute of Technology, Department of Earth, Atmospheric \& Planetary Sciences, Cambridge, MA 02139}
\affiliation{2}{Northwestern University, Center for Interdisciplinary Exploration and Research in Astrophysics \& Department of Physics and Astronomy, Evanston, IL 60601}
\affiliation{3}{Blue Marble Space Institute of Science, Seattle, WA 98154}
\affiliation{4}{Harvard-Smithsonian Center for Astrophysics, Cambridge, MA 02138}
\affiliation{5}{University of Washington, Seattle, WA 98195}
\affiliation{6}{Harvard College, Cambridge, MA 02138}

\affiliation{*}{SCOL Postdoctoral Fellow}





\correspondingauthor{Sukrit Ranjan}{sukrit.ranjan@northwestern.edu}





%
%

%
%


\begin{abstract}
Ultraviolet (UV) light plays a key role in surficial theories of the origin of life, and numerous studies have focused on constraining the atmospheric transmission of UV radiation on early Earth. However, the UV transmission of the natural waters in which origins-of-life chemistry (prebiotic chemistry) is postulated to have occurred is poorly constrained. In this work, we combine laboratory and literature-derived absorption spectra of potential aqueous-phase prebiotic UV absorbers with literature estimates of their concentrations on early Earth to constrain the prebiotic UV environment in marine and terrestrial natural waters, and consider the implications for prebiotic chemistry. We find that prebiotic freshwaters were largely transparent in the UV, contrary to assumptions by some models of prebiotic chemistry. Some waters, e.g., high-salinity waters like carbonate lakes, may be deficient in shortwave ($\leq220$ nm) UV flux. More dramatically, ferrous waters can be strongly UV-shielded, particularly if the Fe$^{2+}$ forms highly UV-absorbent species like Fe(CN)$_6^{4-}$. Such waters may be compelling venues for UV-averse origin-of-life scenarios, but are disfavorable for some UV-dependent prebiotic chemistries. UV light can trigger photochemistry even if attenuated through photochemical transformations of the absorber (e.g., $e^{-}_{aq}$ production from halide irradiation), which may have both constructive and destructive effects for prebiotic syntheses. Prebiotic chemistries invoking waters containing such absorbers must self-consistently account for the chemical effects of these transformations. The speciation and abundance of Fe$^{2+}$ in natural waters on early Earth is a major uncertainty, and should be prioritized for further investigation as it plays a major role in UV transmission in prebiotic natural waters. 
\end{abstract} 


%
%

\section{Introduction\label{sec:intro}}
A key challenge for origin-of-life studies is constraining the range of environmental conditions  on early Earth under which life arose. Knowledge of these environmental conditions informs development of theories of the origin of life, and enables assessment of the plausibility and probability of postulated prebiotic chemistries \citep{Pace1991, Cleaves2013,Barge2018, Lyons2020}. Consequently, considerable work has been invested to constrain the range of physico-chemical conditions available in prebiotic environments \citep{Sleep2018, Sasselov2020}. 

An important prebiotic environmental factor, particularly for surficial prebiotic chemistry, is the ultraviolet (UV) irradiation environment. UV photons can destroy nascent biomolecules, motivating a search for mechanisms to protect prebiotically important molecules from UV irradiation \citep{Sagan1973, Holm1992, Cleaves1998}. On the other hand, UV light has also been suggested to be essential to the origin of life, as a source of chemical free energy and selectivity \citep{Deamer2010, Pascal2012, Beckstead2016}. Indeed, UV light has been experimentally shown to drive a range of prebiotic chemistries \citep{Ferris1966, Sagan1971,  Flores1977, Bonfio2017, Pestunova2005, Mariani2018, Xu2020}. Surficial prebiotic chemistries therefore broadly fall into two classes: those for which UV light is strictly destructive and must be mitigated or avoided, and those for which UV light is essential and must be sought \citep{Ranjan2016, Todd2020}. The self-consistent availability of UV radiation is consequently a crucial part of the assessment of the plausibility of proposed prebiotic chemistries.

The pivotal roles that UV radiation plays in diverse prebiotic chemistries have motivated increasingly sophisticated estimates of the prebiotic UV environment, with particular emphasis on atmospheric transmission. Models generally predict abundant steady-state UV at wavelengths $\gtrsim 204$ nm and a shortwave cutoff of $\approx 200$ nm driven by atmospheric CO$_2$ and H$_2$O, though transient eras of low UV are possible after large volcanic eruptions and possibly impacts \citep{Cockell2000oceans, Rugheimer2015, Ranjan2017a, Ranjan2018, Zahnle2020}. However, prebiotic chemistry is generally proposed to occur in aqueous reservoirs like ponds or oceans, which may host UV-blocking compounds of geological origin \citep{Martin2008, McCollom2013, Patel2015, Benner2019, Becker2019}. To date, estimates of UV transmission in prebiotic natural waters are generally based on pure water or modern pond water \citep{Cockell2000oceans, Ranjan2016, Pearce2017}. 

Neither modern pond water nor pure water are likely representative of prebiotic waters. The UV-opacity of modern pond waters is largely driven by biogenic dissolved carbon species \citep{Morris1995, Markager2000, Laurion2000}. In the limit of low biological productivity, as expected on a prebiotic world, natural waters can be clear down to 300 nm; their transmission from 200-300 nm is unconstrained, but has been extrapolated to be similarly low \citep{Smith1981, Morel2007}.  At the other extreme,  while pure water is clear in the near-UV \citep{Quickenden1980}, the waters in which prebiotic chemistry occurred must by definition have been impure, since they must have contained at minimum the feedstocks for that chemistry, and likely other geogenic constituents as well (e.g., \citealt{Toner2020}). Absorption due to these constituents may influence UV transmission. The UV transmission of prebiotic waters therefore likely lies in between the extremes of high transparency and high opacity represented by the previously-utilized proxies of pure water and modern pond water.

In this paper, we constrain wavelength-dependent UV transmission in prebiotic waters. We focus on absorption due to a subset of UV-active species which have been proposed to be present in prebiotic waters. We derive molar absorptivities of these compounds from a combination of literature reports and our own measurements. We draw on literature proposals for the concentrations of these species in representative prebiotic waters. We do not attempt self-consistent geochemical modelling of these waters; such work is important but requires improved measurements of the relevant reaction kinetics under prebiotically relevant conditions, which is beyond the scope of this work. Our work is similar in spirit to the work of \cite{Cockell2000oceans, Cockell2002}, who drew on literature proposals for the composition of the prebiotic atmosphere to constrain its UV transmission, without attempting to model its self-consistent photochemistry. Their simple initial analysis guided prebiotic chemistry (e.g., \citealt{Pierce2005, Gomez2007}) and motivated more atmospherically sophisticated follow-up work (e.g., \citealt{Rugheimer2015}), which came to largely similar conclusions. Similarly, we hope our simple study will guide prebiotic chemistry while motivating more geochemically sophisticated follow-up work.

\section{Background}
Relatively few studies constrain the UV transmission of prebiotic natural waters specifically. A notable exception is \citet{Cleaves1998}, who considered potential "sunscreens" for the prebiotic ocean. \citet{Cleaves1998} measured the UV absorptivities of salts, HCN and spark-discharge-derived polymers, HS$^-$, and marine Fe$^{2+}$. Based on a combination of calculated and literature estimates of the concentrations of these species, \citet{Cleaves1998} concluded that in favorable circumstances, HS$^-$, marine Fe$^{2+}$, or spark discharge polymer could have extincted UV in the surface layer of the prebiotic ocean. The conditions required for accumulation of spark-discharge polymer to high concentrations in the prebiotic ocean included the emission of prebiotic volcanic carbon as CH$_4$, its conversion with unity efficiency to spark-discharge polymer, and its efficient deposition to the ocean. However, more recent modelling indicates volcanogenic carbon was emitted even in early Earth-like planets as CO$_2$, not CH$_4$, and that the primary fate of abiotic CH$_4$ should have been oxidation to CO$_2$  \citep{Kasting2014}, suggesting the conditions required for accumulation of optically-relevant concentrations of spark discharge polymers were unlikely to be met in the steady state, though high hydrocarbon concentrations may be transiently possible after large impacts \citep{Genda2017, Benner2019, Zahnle2020}. Similarly, more recent work suggests the early ocean was ferrugious (Fe$^{2+}$-rich), and would have titrated out HS$^-$ as pyrite; combined with the low solubility of HS$^-$, [HS$^-$] was most likely very low in most natural waters on early Earth \citep{Walker1985, Poulton2011, Ranjan2018}. However, elevated [Fe$^{2+}$] remains possible for the early ocean \citep{Konhauser2017}.

\section{Methods}
\subsection{Calculating Aqueous UV Attenuation}
We approximate the transmission of UV radiation in homogenous (well-mixed), non-scattering aqueous solutions at low concentrations and low light intensities  by the Beer-Lambert Law \citep{IUPAC_Beer_Lambert}:
\begin{equation}
    \log_{10}(I(\lambda)/I_0(\lambda))=\log_{10}(T(\lambda))=-(\Sigma_i\epsilon_i(\lambda) c_i)d = -(\Sigma_i a_i(\lambda))d,
\end{equation}
where $I_0$ is the incident irradiance, $I$ is the transmitted irradiance, $T$ is the fraction of transmitted radiation, $\epsilon_i$ is the molar decadic absorption coefficient for the $i$th component of the solution, $c_i$ is the concentration of the $i$th component of the solution, $a=\epsilon c$ is the linear decadic absorption coefficient, and $d$ is the path length. This approach considers only absorption and neglects scattering. This is a reasonable approximation because in the $200-300$ nm wavelength range we focus on, the single-scattering albedo of liquid water $\omega_0<<1$, i.e. absorption dominates scattering (Fig.~\ref{fig:waterbeerlambert}; \citealt{Quickenden1980, Krockel2014}). We follow previous workers in assuming that $\epsilon_i$ does not vary as a function of pH for the simple inorganic molecules we consider, i.e. that changes in the absorbance of solutions of these molecules with pH are due to changes in speciation or complexation, not intrinsic changes to $\epsilon_i$ \citep{Braterman1983, Anbar1992, Nie2017, Tabata2021}. We propagate the uncertainties on $\epsilon_i$ under the assumption that they are independent and normally distributed \citep{Bevington2003}. 

\subsection{Molar Decadic Absorption Coefficients for Potential Prebiotic Absorbers}
 We consider absorption due to halide anions (Cl$^-$, Br$^-$, I$^-$), ferrous iron species (Fe$^{2+}$, Fe(CN)$_6^{4-}$), sulfur species (HS$^-$, HSO$_3^-$, SO$_3^{2-}$), dissolved inorganic carbon (HCO$_3^-$, CO$_3^{2-}$) and nitrate (NO$_3^-$). Halides are ubiquitous in natural waters on modern Earth due to their high solubility and robust geological source, and hence are thought to have been present in natural waters on early Earth as well \citep{Knauth2005, Marty2018, hanley2018, Toner2020}. Ferrous iron is inferred on early Earth on the basis of banded iron formations \citep{Walker1985, Li2013, Konhauser2017, Toner2019}. Dissolved inorganic carbon is inevitable in natural waters on early Earth due to dissolution of atmospsheric CO$_2$ \citep{Krissansen-Totton2018ocean, kadoya2020}. Sulfur species may have been present on early Earth due to dissolution of volcanically outgassed sulfur species \citep{Walker1985, Ranjan2018}. Nitrate is predicted to accumulate in natural waters on early Earth as a product of lightning in an N$_2$-CO$_2$ atmosphere \citep{Mancinelli1988, Wong2017, Laneuville2018, Ranjan2019}. 
 
 We draw on both the literature and our own measurements for the molar decadic absorption coefficients of potential prebiotic aqueous-phase absorbers. We use our own measurements when available, because our measurements typically feature broader wavelength coverage than the literature, include estimates of uncertainty, and are collected using uniform techniques (Appendix~\ref{sec:molabs_meas}). For the ferrous iron species, we check that the dominant ion speciation does not vary due to pH drift over the course of our dilutions (Appendix~\ref{sec:concerns}). Our measurement techniques are potentially inaccurate for weak acids and bases because we do not control pH, which may drift during dilution steps, affecting speciation. For such species, we rely instead on literature data (Appendix~\ref{sec:molabs_lit}). Table~\ref{tbl:molar_decadic_absorption_coefficients} summarizes the sources of the molar decadic absorption coefficients used in this work. 

\begin{table}[h]
\caption{Molar decadic absorption coefficients used in this work\label{tbl:molar_decadic_absorption_coefficients}}
\centering
\begin{tabular}{p{2.6cm}p{7.7cm}}
\hline
Species  &  Source \\
\hline
Br$^-$  &  As NaBr, this work\\
Cl$^-$  &  As NaCl, this work\\
I$^-$  &  As NaI and KI, this work\\
NO$_3^-$  &  As NaNO$_3$, this work\\
Fe$^{2+}$  &  As Fe(BF$_{4}$)$_2$, this work\\
Fe(CN)$_6^{4-}$  &  As K$_4$Fe(CN)$_6$, this work\\
HS$^-$  &  \cite{Guenther2001}\\
HSO$_3^-$ &  \cite{Fischer1996, Beyad2014}\\
SO$_3^{2-}$ &  \cite{Fischer1996, Beyad2014}\\
HCO$_3^-$ &  \cite{Birkmann2018}\\
CO$_3^{2-}$ &  \cite{Birkmann2018}\\
H$_2$O  &  \cite{Quickenden1980}\\
\\
\hline
\end{tabular}
\end{table}

\subsection{Abundances of Potential Prebiotic Absorbers in Fiducial Prebiotic Waters}
Constraining the impact of potential absorbers on prebiotic aqueous transmission requires estimates of their abundances in natural waters on early Earth. Natural waters are diverse, and we cannot hope to explore this full diversity. Instead, we focus on fiducial waters motivated by proposed origin-of-life scenarios. We focus primarily on shallow terrestrial waters, e.g. ponds and lakes. Such waters are of interest for prebiotic chemistry because of their propensity for wet-dry cycles, the ability to accumulate atmospherically-delivered feedstocks more efficiently than the oceans, and their potentially diverse palette of environmental conditions \citep{Patel2015, Deamer2017, Pearce2017, Becker2018, Rimmer2019, Ranjan2019, Toner2020}. We specifically consider freshwater lakes, carbonate lakes, and ferrocyanide lakes. We also consider the early ocean, to offer a basis of comparison for the terrestrial waters. Our oceanic calculations may also be relevant to origin-of-life scenarios that invoke shallow waters at the land-ocean interface \citep{Commeyras2002, Lathe2005, Bywater2005}, but are not relevant to deep-sea origin of life scenarios (e.g., \citealt{Corliss1981, Sojo2016}), where water alone is enough to extinct UV. The composition of these waters are uncertain: we draw on the literature to construct high and low transmission endmember cases, to bound their potential UV transmission. Our construction of these endmember cases is summarized in Table~\ref{tbl:fiducial_prebiotic_waters}, and detailed in Sections~\ref{sec:ocean_comp}-\ref{sec:ferrouslake_comp}. 

\begin{table}[h]
\caption{Estimated range of concentrations of potential prebiotic absorbers in prebiotic waters\label{tbl:fiducial_prebiotic_waters}}
\centering
\begin{tabular}{l|ll|ll|ll|ll}
\hline
Species  &  Ocean & & Freshwater  & Pond &  Carbonate  & Lake& Ferrous  & Lake\\
& Low & High & Low & High & Low& High & Low & High \\
\hline
Br$^-$  &  0.45 mM &  1.8 mM & 0.15 $\mu$M& 0.15 $\mu$M & 1 mM& 10 mM & 0.15 $\mu$M & 0.15 $\mu$M \\
Cl$^-$  &  0.3 M &  1.2 M & 0.2 mM & 0.2 mM & 0.1 M & 6M & 0.2 mM & 0.2 mM\\
I$^-$  &  0.25 $\mu$M &  1 $\mu$M  & 40 nM & 600 nM & 40 nM & 600 nM & 40 nM & 600 nM \\
Fe$^{2+}$  &  1 nM &  0.1 mM  & 0.1 $\mu$M & 0.1 mM & 0 & 0 & 0 & 0\\
Fe(CN)$_6^{4-}$  &  0 &  0  & 0 & 0 & 0 & 0 & 0.1 $\mu$M & 0.1 mM\\
NO$_3^-$  &  $2\times10^{-15}$M &  0.5 $\mu$M  & 0.05 nM & 10 $\mu$M & 5 nM & 1 mM & 0.05 nM& 10 $\mu$M\\
HS$^-$  &  0 &  0  & 0 & $8\times10^{-11}$M & 0 & 8 nM  & 0 & $8\times10^{-11}$M \\
HSO$_3^-$  &  0 &  0  & 0 & 200 $\mu$M & 0 & 200 $\mu$M & 0 & 200 $\mu$M \\
SO$_3^{2-}$  &  0 &  0  & 0  & 100 $\mu$M  & 0 & 400 $\mu$M & 0 & 100 $\mu$M \\
HCO$_3^-$  &  2 mM &  0.2M  & 1 mM & 1 mM & 50 mM & 0.1 M & 1 mM & 1 mM \\
CO$_3^{2-}$  &  0.2 $\mu$M &  1 mM  & 100 nM & 100 nM & 7 $\mu$M & 7 mM &  100 nM &  100 nM \\
\\
\hline
\end{tabular}
\end{table}

\subsubsection{Ocean\label{sec:ocean_comp}}
For halide species, we scale the composition of the modern oceans. We consider a salinity range of $0.5-2\times$ modern, motivated by theoretical arguments and isotopic evidence \citep{Knauth2005, Knauth1998, Marty2018}. On modern Earth, seawater halide concentrations are [Cl$^-$]=0.6M, [Br$^-$]=0.9 mM, and [I$^-$]=0.5 $\mu$M \citep{Channer1997, ASTM_SeaWater}. Cl$^-$ and Br$^-$ covary in natural waters, leading us to fix their prebiotic ratios to the modern value \citep{hanley2018}. \citet{deRonde1997} used fluid inclusions to infer high [I$^-$] at 3.2 Ga, but the interpretation of these samples is strongly contested \citep{Lowe2003, Knauth2005, Farber2015}. We therefore fix our prebiotic [I$^-$]/[Cl$^-$] to the modern value as well, and assume these ions to covary in the prebiotic ocean. 

Estimates of ferrous iron concentrations in the surficial prebiotic ocean vary significantly. \citet{Halevy2017greenrust} estimate [Fe$^{2+}$]$<10^{-9}$M in the surface ocean, largely driven by the assumption of efficient photooxidation. More recent work reports this process to be less efficient and estimates [Fe$^{2+}$]$=10^{-4}$M in the surface ocean \citep{Konhauser2007, Halevy2017, Konhauser2017}. Fe$^{2+}$ is predicted to be the main ferrous species at circumneutral pH for solutions with oceanic Cl$^-$ and SO$_4^{2-}$ concentrations in equilibrium with 0.03 atm CO$_2$, which approximates conditions for the Archaean ocean \citep{King1998, Halevy2017, Krissansen-Totton2018ocean}. We therefore assume the ferrous iron to be present as Fe$^{2+}$, and we adopt $10^{-9}$M and $10^{-4}$M as our bracketing estimates for [Fe$^{2+}$] in the photic zone.

We take the bracketing range of oceanic [NO$_3^-$] from \citet{Ranjan2019}. We take oceanic sulfide concentrations to be negligible due to titration with Fe$^{2+}$ \citep{Walker1985, Poulton2011}. We take oceanic sulfite and bisulfite concentrations to be negligible \citep{Halevy2013}. We take bracketing oceanic carbonate and bicarbonate concentrations spanning the range at 3.9 Ga defined by "standard" and "control" cases of the early ocean model of \citet{kadoya2020}.

\subsubsection{Freshwater Lakes \label{sec:freshwaterlake_comp}}
Terrestrial waters are typically dilute, with ionic strength of order $10^{-3}$ M \citep{PhysChemLakes}. To represent a a dilute endmember scenario for natural waters, we consider freshwater lakes with riverine composition (i.e., not evaporatively concentrated). Modern surface freshwater systems average [Cl$^-$]=0.2 mM, and the same is proposed for Archaean and prebiotic river waters \citep{Graedel1996, Hao2017}. Mean [Br$^-$]/[Cl$^-$]$\approx1-2\times10^{-3}$ in modern terrestrial waters \citep{Edmunds1996, Magazinovic2004}, suggesting mean [Br$^-=0.1-0.2\mu$M]. Mean iodide concentrations in modern terrestrial waters is reported as $40$ nM (range: $0.1$nM - $0.6\mu$M) \citep{Fuge1986}, and it is not clear that I$^-$ is generally correlated with Cl$^-$ as Br$^-$ seems to be (e.g., \citealt{Worden1996}). We therefore consider both mean-I (40 nM) and high-I (0.6 $\mu$M) compositions. Estimates of [Fe$^{2+}$] in riverine waters on early Earth span $0.1 \mu$M- $0.1$ mM \citep{Halevy2017greenrust, Hao2017}, and we consider this range. \citet{Hao2017} predict Archaean river water to have pH$\leq6.34$, for which Fe$^{2+}$ should mainly present as Fe$^{2+}_{aq}$ \citep{King1998, Tabata2021}. We take [HCO$_3^{-}$] from \citet{Hao2017}, and calculate [CO$_3^{2-}$] from it. We take the bracketing range of [NO$_3^-$] from \citet{Ranjan2019}, scaled by $0.01\times$ to remove the assumption of a high drainage ratio.  We take upper bounds on [SO$_3^{2-}$], [HSO$_3^-$] and [HS$^-$] from \citet{Ranjan2018} for circumneutral pH and steady-state outgassing. We consider a lower bound of 0 for all sulfur species following the assumption of \citet{Halevy2013} for rivers. 

\subsubsection{Closed-Basin Carbonate Lake\label{sec:carbonatelake_comp}}
Closed-basin carbonate lakes have been proposed as venues for prebiotic chemistry, because the elevated carbonate concentrations in these lakes suppresses [Ca$^{2+}$], permitting the accumulation of phosphate to prebiotically-relevant concentrations \citep{Toner2020}. For a representatative endmember, we consider a closed basin carbonate lake with $10^{-2}$ mol/kg phosphorus, corresponding to the upper edge of phosphorus concentrations in the sample of closed-basin carbonate lakes reported by \citet{Toner2020}. Cl$^-$ and Br$^-$ behave conservatively with P in carbonate lakes, with [Cl$^-$]$=0.1-6$ mol/kg and [Br$^-$] $=10^{-3}-10^{-2}$ mol/kg for [P]$=10^{-2}$ mol/kg  \citep{Toner2020}. In projecting from \citet{Toner2020}, we simplify the calculation by approximating molarity as molality; we contend this approximation to suffice for the order-of-magnitude estimates we seek. Iodine is proposed to be delivered to surface waters via rainfall, and so we might expect it to be evaporatively concentrated in closed-basin lakes; however, we located no reports of evaporative iodide concentration in composition studies of such lakes \citep{Fuge1986, Toner2020, Hirst1995, Mochizuki2018,Eugster1979, Friedman1976}. We therefore consider a bracketing range of [I$^-$] equal to the freshwater lake scenario. We take Fe$^{2+}_{aq}=0$, as high carbonate concentrations should suppress Fe$^{2+}_{aq}$ due to siderite precipitation \citep{Toner2020}. \citet{Toner2020} do not report [HCO$_3^-$] or [CO$_3^{2-}$] for closed-basin carbonate lakes on early Earth, but do calculate that a bracketing range of pCO$_2$=(0.01 bar, 1 bar) to correspond to pH=(9, 6.5) in such lakes. We convert these estimates to approximate bracketing ranges on [HCO$_3^-$] and [CO$_3^{2-}$] under assumption of equilibrium \citep{Sander2015,CRC98}. We take [NO$_3^-$] from \citet{Ranjan2019}. We consider a lower bound of 0 for all sulfur species following the assumption of \citet{Halevy2013} for riverine waters. We take upper bounds on [HS$^-$], [SO$_3^{2-}$] and [HSO$_3^-$] from \citet{Ranjan2018} for steady-state outgassing, for pH=7 and pH=9. The upper limits listed on [SO$_3^{2-}$] and [HSO$_3^-$] cannot simultaneously be achieved, because they correspond to different pHs. Our conclusions are robust to this inconsistency, because even at their respective upper limits SO$_3^{2-}$ and HSO$_3^-$ are not the dominant absorbers in the carbonate lake scenario.

\subsubsection{Ferrous Lakes \label{sec:ferrouslake_comp}}
Ferrocyanide lakes have been proposed to form when ferrocyanide salt deposits are irrigated by neutral water \citep{Toner2019, Sasselov2020}. Ferrocyanide is an extremely potent UV absorber, motivating us to consider transmission in such ferrocyanide lakes. Notably, ferrocyanide has been invoked in UV-dependent prebiotic chemistries, providing an opportunity to check their geochemical self-consistency \citep{Xu2018, Mariani2018, rimmer2018origin}. We approximate the composition of a ferrocyanide lake as the freshwater lake, but with the Fe$^{2+}$ present as Fe(CN)$_6^{4-}$ instead.

\section{Results}

\subsection{Prebiotic Ocean\label{sec:prebiotic-ocean}}

\begin{figure}[h]
\noindent\includegraphics[width=1\textwidth]{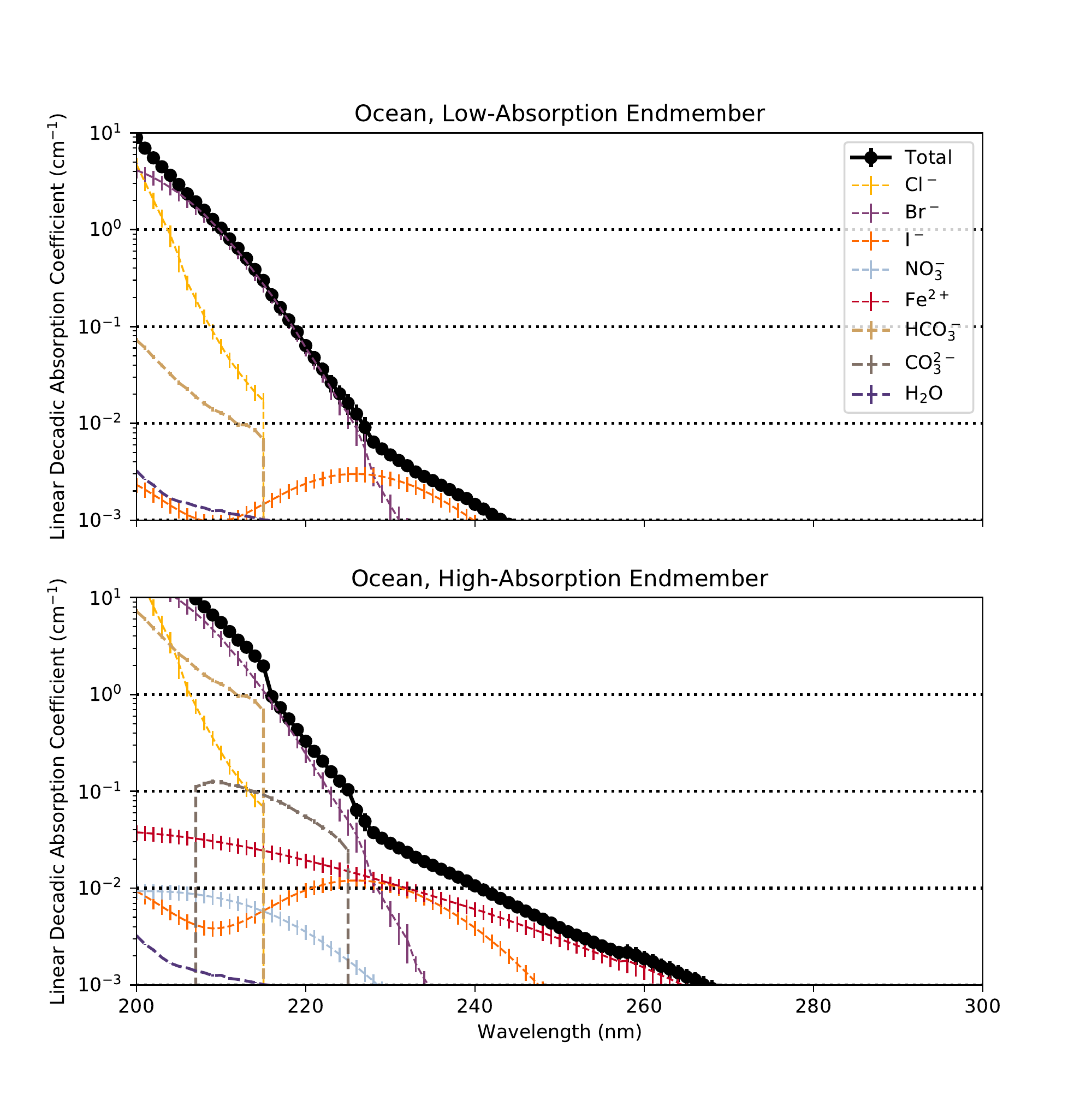}
\caption{Simulated linear decadic absorption coefficients of the prebiotic ocean and its component solutes, for low-absorption and high-absorption endmember cases. Not all solutes are visible in each case, because the linear decadic absorption coefficients of some solutes fall below the lower limit of the y-axis across the wavelength space plotted here. The shortest-wavelength photons are extincted in the surface layers of the ocean, but the  $\sim260$ nm radiation responsible for nucleotide degradation can penetrate to a depth of meters.}
\label{fig:halide-ferrous-ocean}
\end{figure}

The prebiotic ocean efficiently attenuated shortwave UV radiation, but may have admitted longer-wavelength UV radiation to depths of meters (Figure~\ref{fig:halide-ferrous-ocean}). In the low-absorption endmember, the absorption is dominated by the halides, especially Br$^-$ at shorter wavelengths and I$^-$ at longer wavelengths. In the high-absorption endmember scenario, absorption is dominated by Br$^-$ at shorter wavelengths and Fe$^{2+}_{aq}$ at longer wavelengths. Halides confine the shortest-wavelength UV photons to the surface of the ocean: in even the low-absorption endmember scenario, the ocean is optically thick\footnote{Incident flux attenuated by $\geq\frac{1}{e}$ \citep{Thomas2002}} at $ \leq 220$ nm  for depths $d\geq7\pm 1$ cm, driven primarily by Br$^-$. Longer-wavelength radiation may penetrate to much greater depths, with the absorbers considered here permitting penetration of the $\sim260$ nm radiation responsible for nucleotide degradation to a depth of meters in even the high-absorption endmember scenario.

\subsection{Prebiotic Terrestrial Waters}

\subsubsection{Freshwater Lakes}

\begin{figure}[h]
\noindent\includegraphics[width=1\textwidth]{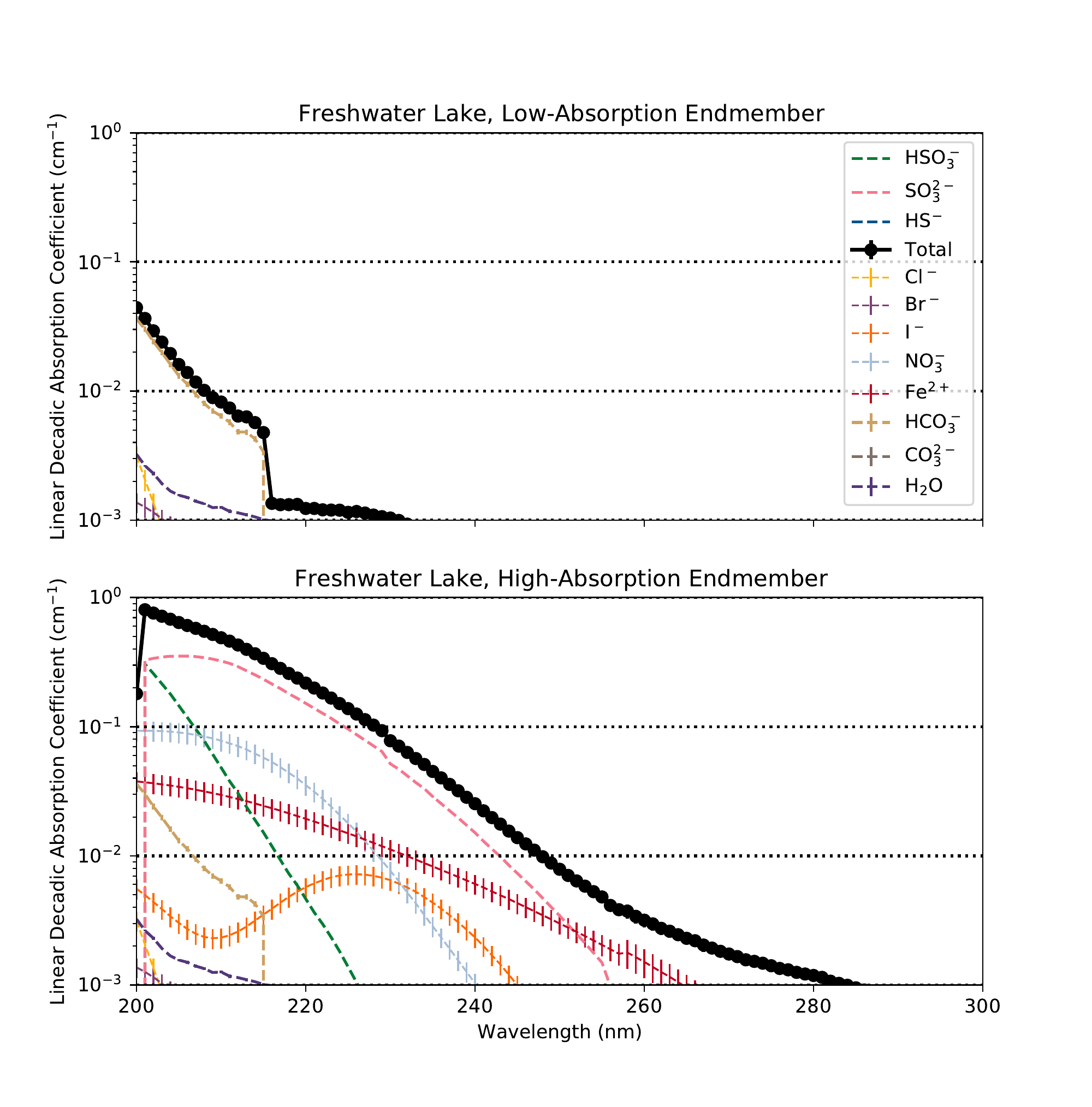}
\caption{Simulated linear decadic absorption coefficients of the prebiotic freshwater lake and its component solutes, for low-absorption and high-absorption endmember cases. Not all solutes are visible in each case, because the linear decadic absorption coefficients of some solutes fall below the lower limit of the y-axis across the wavelength space plotted here. The freshwater lake may have been largely transparent to UV, and even in the high-absorption endmember would have been largely transparent to UV at the longer wavelengths ($\sim260$ nm) relevant to nucleotide photolysis.}
\label{fig:halide-freshwater-lake}
\end{figure}

The transparency of shallow freshwater lakes depends strongly on the abundances of Fe$^{2+}$ and S(IV) species (sulfite, bisulfite). In the low-absorption endmember scenario, freshwater lakes are essentially transparent in the UV, with depths of meters required for non-negligible attenuation of UV across most of the UV (Figure~\ref{fig:halide-freshwater-lake}). However, in the high-absorption endmember scenario, shortwave UV is shielded due to sulfite, with secondary shielding from nitrate and Fe$^{2+}_{aq}$. For radiation with wavelengths $\leq 235$ nm, the high-absorption endmember freshwater lake is optically thick for depths $d\geq9.6\pm0.4$ cm. However, even the highly absorptive endmember remains optically thin to the $\gtrsim260$ nm radiation which dominates nucleotide photolysis down to a depth of $1.4\pm 0.2$ m. 

\subsubsection{Closed-Basin Carbonate Lakes \label{sec:carbonate-lakes}}
\begin{figure}[h]
\noindent\includegraphics[width=1\textwidth]{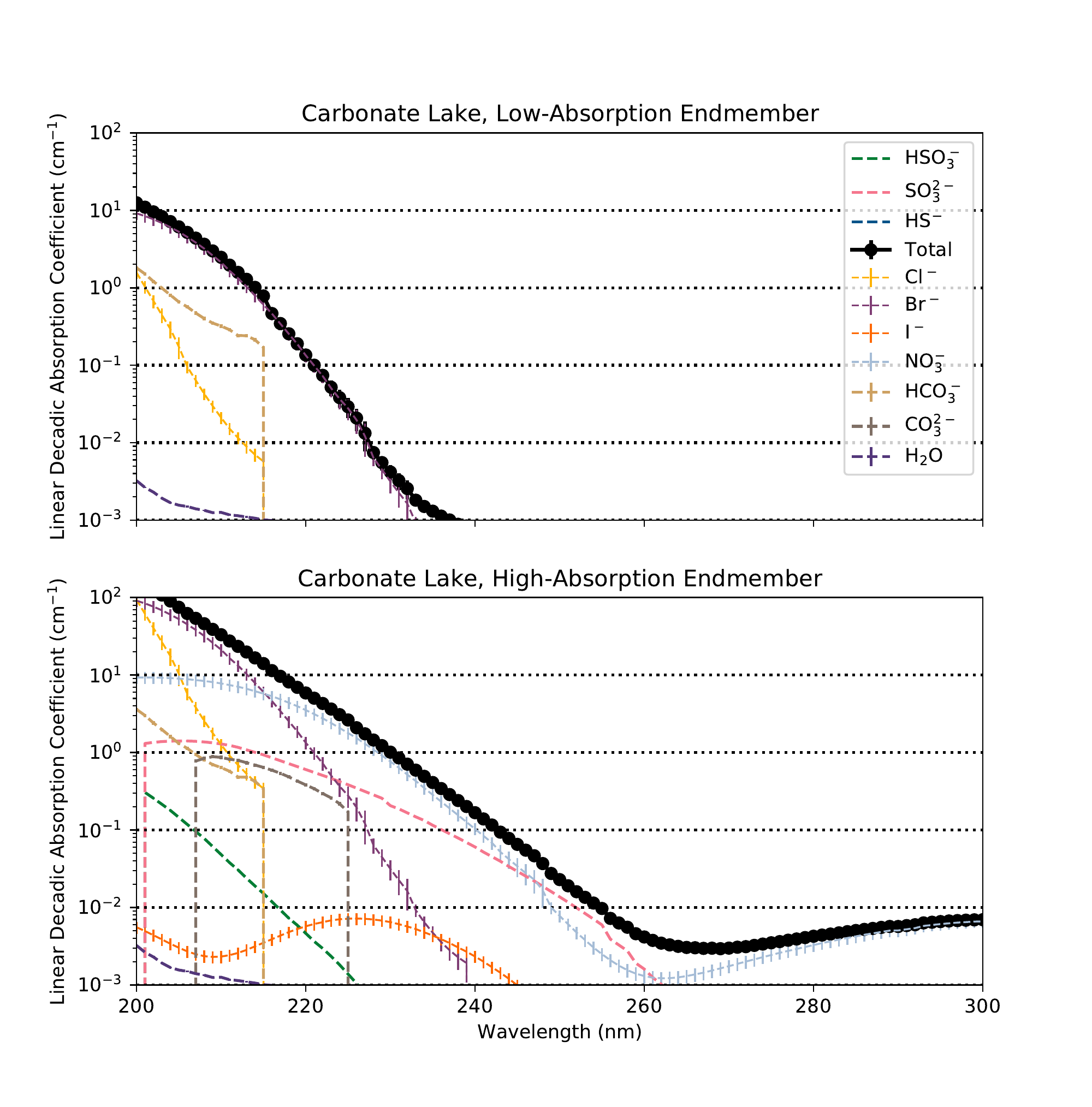}
\caption{Simulated linear decadic absorption coefficients of the prebiotic carbonate lake and its component solutes, for low-absorption and high-absorption endmember scenario. Not all solutes are visible in each case, because the linear decadic absorption coefficients of some solutes fall below the lower limit of the y-axis across the wavelength space plotted here. The carbonate lake robustly extincts shortwave UV in even the low-absorption endmember scenario, driven by Br$^-$. In the high-absorption endmember case, longer wavelength UV would have been available throughout shallow lakes ($d<1$m), but would have been extincted at depths of a few meters.}
\label{fig:halide-carbonate-lake}
\end{figure}

Closed-basin carbonate lakes can be more UV-opaque compared to the prebiotic ocean (Figure~\ref{fig:halide-carbonate-lake}). Elevated [Br$^-$] robustly limits absorption at short wavelengths; in even the low-absorption endmember scenario,  $\leq220$ nm radiation is efficiently extincted (i.e., is in the optically thick regime) for depths of $>3.2\pm 0.6$ cm. Radiation at $260$ nm is robustly available for depths $\leq 1.04\pm 0.07$ m even in the high absorption endmember scenario, but is depleted at depths of a few meters, driven by nitrate absorption with contributions from sulfite. 

\subsection{Ferrocyanide Lakes}
 
\begin{figure}[h]
\noindent\includegraphics[width=1\textwidth]{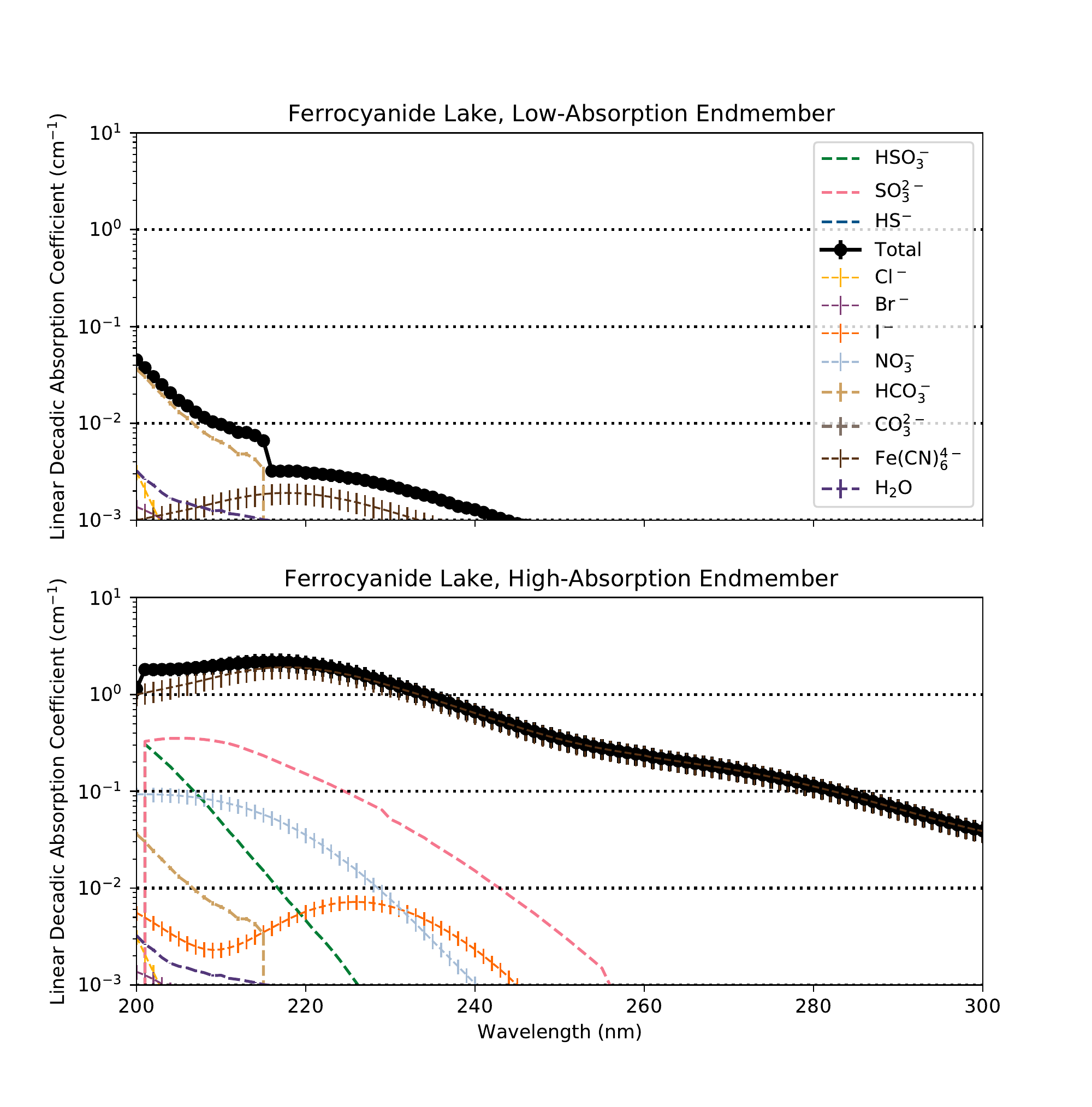}
\caption{Simulated linear decadic absorption coefficients of the prebiotic ferrocyanide lake and its component solutes, for low-absorption and high-absorption endmember scenarios. Not all solutes are visible in each case, because the linear decadic absorption coefficients of some solutes fall below the lower limit of the y-axis across the wavelength space plotted here.  Ferrocyanide is an effective sunscreen, and lakes hosting more than dilute ferrocyanide ($\gtrsim100~\mu$M) would have been low-UV environments.}
\label{fig:ferrous-lake}
\end{figure}

Ferrocyanide lakes may have been low-UV environments (Figure~\ref{fig:ferrous-lake}). Ferrocyanide is a stronger, and more crucially, broader UV absorber than Fe$^{2+}$, able to attenuate UV across the near-UV range relevant to prebiotic chemistry. Attenuation due to ferrocyanide is minimal in the low-absorption endmember scenario. However, for the high absorption endmember, the ferrocyanide lake becomes optically thick by $d=11\pm3$ cm across 200-300 nm. Photochemical derivatives of ferrocyanide, such as nitroprusside and ferricyanide, are similarly potent UV absorbers, suggesting ferrocyanide-rich lakes will remain UV-poor even if some of the ferrocyanide is photochemically processed into these forms \citep{Xu2018, Mariani2018, Ross2018, Strizhakov2014}. Ferrocyanide and derived compounds may therefore have been strong "sunscreens” in select lacustrine environments on early Earth, if present at elevated concentrations as postulated by some authors and invoked by others \citep{Xu2018, Toner2019, Sasselov2020}.
 
\section{Discussion}
In \citet{Ranjan2016}, we argued that UV radiation down to wavelengths of $\sim204$ nm would have been available for aqueous prebiotic chemistry on early Earth, motivated by the transmission of the atmosphere and of pure water. However, prebiotic waters were likely not pure. In this paper, we re-examine the conclusion of \citet{Ranjan2016} by considering potential UV absorbers that might have been present in prebiotic waters.

\subsection{Terrestrial Freshwater Systems May Have Been UV-Transparent}
Prebiotic terrestrial freshwaters may have been largely transparent in the UV. In the low-absorption endmember scenario for terrestrial freshwater lakes, shallow lakes could have been transparent down to depths of meters across most of the $>$200 nm wavelength range admitted by the prebiotic atmosphere, meaning that UV radiation may have been a pervasive aspect of the surficial prebiotic milieu. Some works assume UV-opaque prebiotic pond waters, based on extrapolation of modern natural waters; however, the opacity of modern terrestrial waters in the UV is generally due to biogenic dissolved organic compounds (e.g., \citealt{Morris1995, Markager2000, Laurion2000, Pearce2017}). Indeed, even on modern Earth, waters with low biological productivity are transparent to a depth of $\geq3$ m down to the 300 nm threshold to which their transmission has been characterized \citep{Smith1981, Morel2007}. Prebiotic freshwaters may have been similarly transparent, unless abiotic processes were comparably efficient to modern productive ecosystems in generating organics (e.g., potentially in the immediate aftermath of a large impact; \citealt{Zahnle2020}).

\subsection{Shortwave UV Was Likely Attenuated in Diverse Prebiotic Waters}

\begin{figure}[h]
\noindent\includegraphics[width=0.75\textwidth]{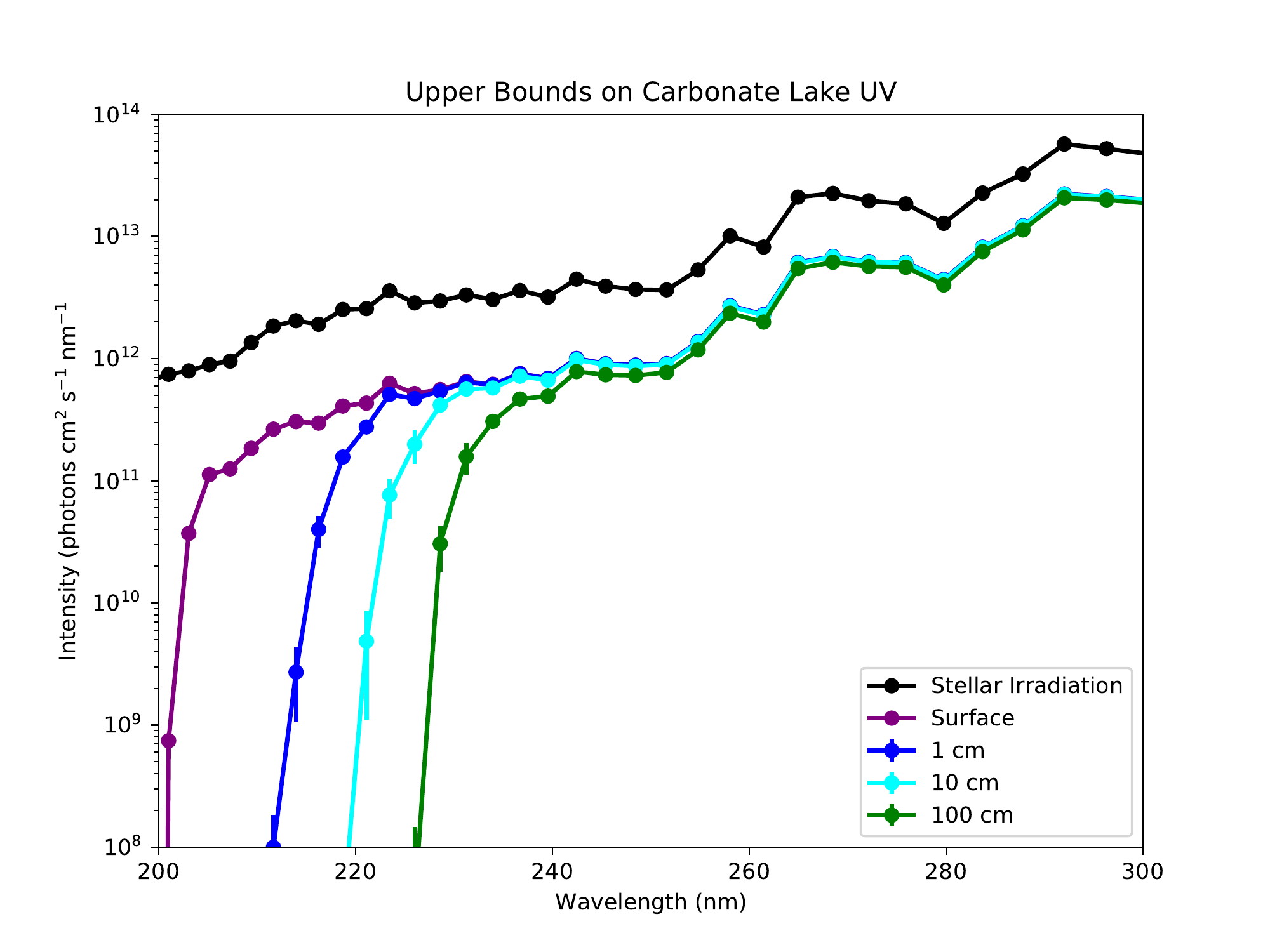}
\caption{Estimate of actinic UV flux as a function of depth in the low-absorption endmember of the carbonate lake scenario, calculated by attenuating the surface actinic flux of \citet{Ranjan2017a} (their "surface radiance") by Beer-Lambert law, assuming a slant angle of $60^\circ$, atmospheric composition from \citet{Rugheimer2015}, and stellar irradiation of 3.9 Ga Sun from \citet{Claire2012}. In even the low-absorption endmember, attenuation of $<220$ nm irradiation is significant.}
\label{fig:uv-halide-carbonate-lake}
\end{figure}

While prebiotic natural waters in general were not necessarily opaque across the UV ($200-300$ nm), shortwave UV ($\leq220$ nm) was attenuated in diverse prebiotic waters. Numerous prebiotic absorbers strongly attenuate shortwave UV, even if they absorb longwave UV only weakly. In particular, in high-salinity waters like the prebiotic ocean and carbonate lakes, the halide anions, especially Br$^-$, efficiently attenuate shortwave UV. In even the low-absorption endmembers, $\lambda\leq 220$ nm radiation is restricted to depths $d\leq7\pm1$ cm in the prebiotic ocean, and $d\leq3.2\pm0.6$ cm in the carbonate lake scenario (Figure~\ref{fig:uv-halide-carbonate-lake}). Shortwave UV may have been even more strongly attenuated in some waters: for example, in the high-absorption endmember for the carbonate lake scenario, nitrate absorption restricts $\lambda\leq 230$ nm UV to $d\leq0.43\pm0.06$ cm. Such nitrate concentrations would only have been available in shallow closed-basin lakes with large drainage ratios, and only if atmospheric NO$_X^-$ production rates were at the upper edge of the predicted range \citep{Ranjan2019}. Shortwave UV may have been restricted in prebiotic terrestrial waters in general, if prebiotic terrestrial sulfite concentrations were at the upper edge of what has been proposed in the literature (Figure~\ref{fig:halide-freshwater-lake}). However, these literature estimates of prebiotic terrestrial sulfite concentrations consider only thermal processes and neglect photolytic loss mechanisms, meaning they be overestimates; further modelling including sulfite photolysis is required to rule on this possibility \citep{Deister1990, Fischer1996, Halevy2013, Ranjan2018}. In sum, shortwave UV may have been strongly attenuated in diverse prebiotic waters, including saline lakes and closed-basin lakes, due to the action of species like Br$^-$ and nitrate. 

\subsection{Broadband UV Was Attenuated in Some Prebiotic Waters}
\begin{figure}[h]
\noindent\includegraphics[width=0.75\textwidth]{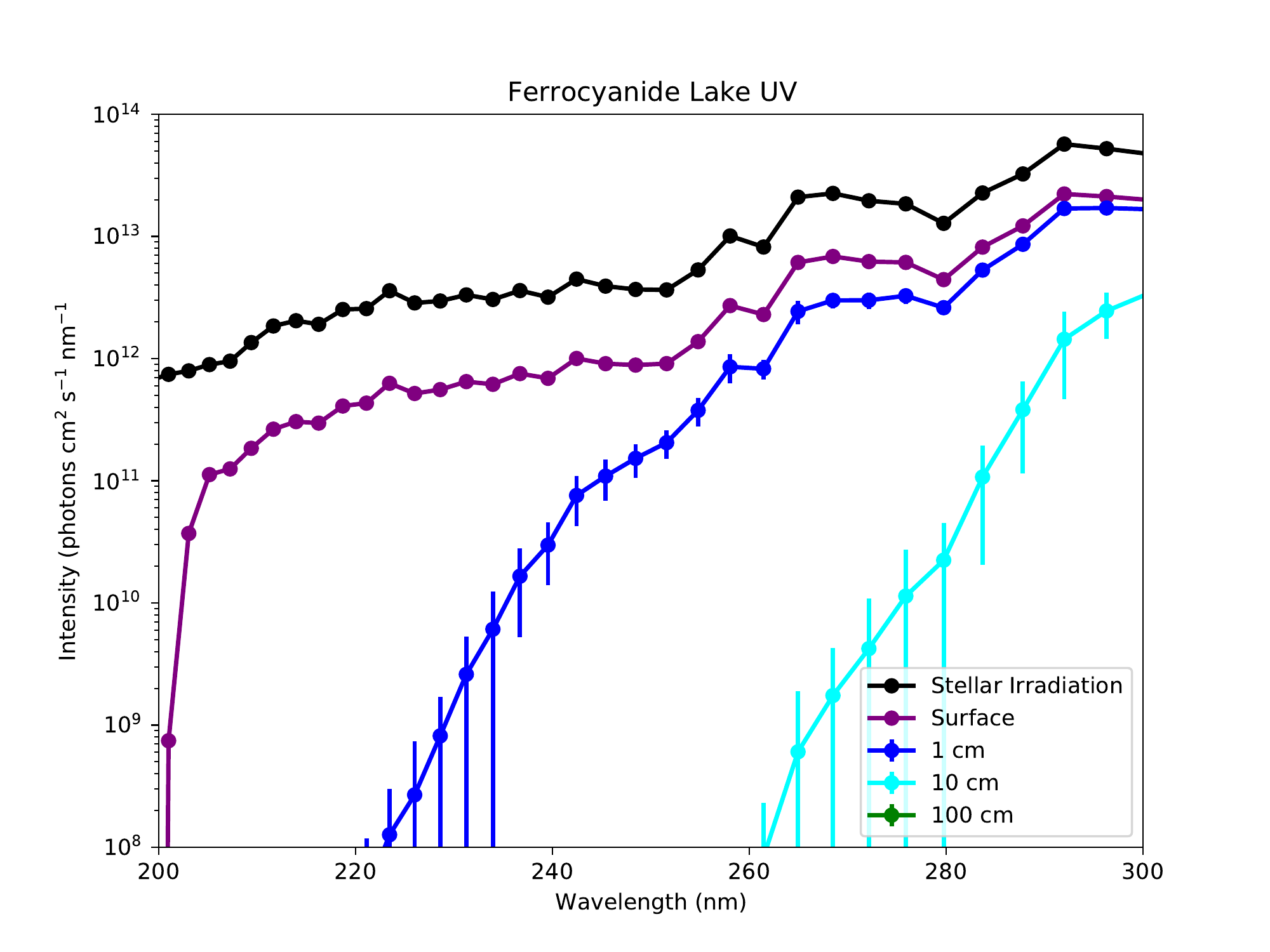}
\caption{Estimate of actinic UV flux as a function of depth in the high-absorption endmember of the ferrocyanide lake scenario, calculated by attenuating the surface actinic flux of \citet{Ranjan2017a} (their "surface radiance") by Beer-Lambert law, assuming a slant angle of $60^\circ$, atmospheric composition from \citet{Rugheimer2015}, and stellar irradiation of 3.9 Ga Sun from \citet{Claire2012}. In the high-absorption endmember, broadband UV is extincted at shallow depths. 
}
\label{fig:uv-ferrous-lake}
\end{figure}

While diverse prebiotically-plausible absorbers are capable of attenuating shortwave UV, fewer prebiotic absorbers are capable of broadband UV attenuation, including the longer-wavelength UV photons which dominate the early Sun's UV output and hence prebiotic photoprocesses like nucleobase photolysis. One prebiotically-proposed family of broadband UV absorbers are compounds derived from ferrous iron. In particular, ferrocyanide is a strong, broad UV absorber.  In the high-absorption endmember scenario for ferrocyanide lakes, ferrocyanide would have suppressed (optical depth $>$1) $\leq 260$ nm radiation for $d\geq1.8\pm0.5$ cm, and $\leq 300$ nm radiation for $d\geq11\pm3$ cm (Figure~\ref{fig:uv-ferrous-lake}). Photochemical derivatives of ferrocyanide, such as ferricyanide and nitroprusside, are similarly effective broadband UV screens (Figure~\ref{fig:SNP_etc}). Ferrocyanide lakes and their derivative waters would have been low-UV environments at depth on early Earth.

The existence and prevalence of ferrocyanide lakes on early Earth is uncertain. In particular, even if ferrocyanide lakes form as proposed by \citet{Toner2019}, ferrocyanide undergoes photoaquation under irradiation by $\leq400$ nm radiation \citep{Asperger1952}. \citet{Toner2019} argue that rapid back-reaction stabilizes ferrocyanide against this photodecomposition, based on the experimental study of \citet{Asperger1952}. However, the measurements of \citet{Asperger1952} were not conducted in prebiotically-representative conditions. In particular, their UV irradiation was not representative of prebiotic UV irradiation, and the ferrocyanide concentrations used in their study were extremely high ($\geq 50$ mM). It is not clear whether ferrocyanide lakes would remain stable under more prebiotically-representative conditions; detailed measurements and modelling of ferrocyanide photochemistry in prebiotically-relevant conditions are required. Alternately, detection of remnants of ferrocyanide-rich environments on Mars (e.g., ferrocyanide salt deposits) might inform our understanding of the prevalence such systems on early Earth \citep{Sasselov2020, Mojarro2021}. 

In addition to ferrocyanide, other ferrous iron compounds like FeSO$_4$ and FeCl$_2$ may also have acted as prebiotic "sunscreens", thanks to their broadband UV absorption; however, much higher concentrations of these compounds are required compared to ferrocyanide due to their much lower molar absorption in the UV (Figure~\ref{fig:SNP_etc}). Additionally, their formation is not thermodynamically favored relative to Fe$^{2+}$ under conditions relevant to natural waters on modern Earth \citep{King1998}, and may not have been favored on early Earth either; modelling under prebiotic Earth conditions is required to definitively rule on this question. Finally, the abundance of Fe$^{2+}$ itself in natural waters on early Earth remains debated \citep{Konhauser2017, Halevy2017greenrust, Hao2017}. Detailed, focused modelling of Fe$^{2+}$ abundances and speciation in prebiotic waters on early Earth is required to estimate the prevalence of Fe$^{2+}$-derived "sunscreens" such as ferrocyanide in natural waters on early Earth

We do not recover the finding of \citet{Cleaves1998} that Fe$^{2+}$ in isolation would be an effective broadband sunscreen at concentrations corresponding to the prebiotic ocean. We report a molar absorbance for the Fe$^{2+}$ ion (as Fe(BF$_4$)$_2$; Section~\ref{sec:fe2-in-isolation}) at 260 nm ($15\pm4$ M$^{-1}$ cm$^{-1}$) that is approximately consistent with \citet{Fontana2007} and \citet{Heinrich1990} but two orders of magnitude lower than \citet{Cleaves1998} (1630 M$^{-1}$ cm$^{-1}$). The molar absorbances of FeCl$_2$, FeSO$_4$, and FeOH$^+$ at 260 nm are also significantly below that reported by \citet{Cleaves1998} \citep{Anbar1992, Fontana2007}. We speculate that one possible explanation might be formation of Fe$^{2+}$ complexes in the study of \citet{Cleaves1998}. At slightly basic pH, characteristic of the modern ocean conditions simulated by \citet{Cleaves1998}, Fe$^{2+}$ may have complexed with OH$^-$ and/or CO$_3^{2-}$ \citep{King1998}, and the absorption properties of such complexes may differ from Fe$^{2+}_{aq}$. Our findings for the prebiotic ocean are more consistent with those of \citet{Anbar1992} for the Archaean ocean, who reported efficient attenuation of shortwave UV in the early ocean but the possibility of transmittance of longer-wavelength UV to a depth of meters.

\subsection{Implications for Prebiotic Chemistry}
Of the species considered in this study, we identify no ubiquitous absorber capable of attenuating broadband UV (200-300 nm) in prebiotic natural waters. Indeed, prebiotic freshwaters may have been essentially transparent in the UV. Prebiotic chemistries which are adversely affected by UV irradiation (UV-avoidant) must either be demonstrated to be so productive as to outpace UV degradation under prebiotic conditions, or invoke a UV-shielded prebiotic milieu (e.g., the organic-haze shielded aftermath of a large impact; \citealt{Benner2019}). UV light in the $\sim240-300$ nm wavelength range was not significantly attenuated by the geologically-derived solutes considered in this study in most terrestrial prebiotic waters, meaning that low-pressure mercury lamps with primary emission at 254 nm remain reasonable proxies for UV flux in initial simulations of prebiotic chemistry. However, follow-up studies (e.g., characterization of action spectra) are required to verify that pathways discovered with such sources could have functioned in realistic natural environments on prebiotic Earth \citep{Ranjan2016, Todd2018, Rimmer2021}.

Diverse prebiotic absorbers are capable of attenuating shortwave UV ($\leq 220$ nm). The general prebiotic importance of removal of  shortwave UV ($\leq 220$ nm) should not be overstated, because most of the young Sun's UV flux was delivered at wavelengths $>220$ nm \citep{Claire2012}. In particular, shallow saline waters do not shield the canonical nucleobases, nucleosides, and nucleotides, whose photolysis is expected to be dominated by the longwave ($\sim260$ nm) bands \citep{Voet1963, Todd2020}. Similarly, the photolysis of 2-aminothiazole and the photoconversion of cytidine to uridine are not inhibited in saline waters \citep{Todd2019, Todd2020}. However, the photolyses of 2-aminooxazole and 2-aminoimidazole, intermediates with proposed roles in prebiotic nucleotide synthesis and activation \citep{Powner2009,Patel2015, Li2017}, are dominated by $<230$ nm radiation under early Earth conditions and their photodestruction lifetimes are hence modestly enhanced in highly saline waters. \citet{Todd2019} estimate half-lives of these molecules to photodestruction to be $\approx$ 7, 26, and 99 hours on early Earth; the half-lives of these molecules to direct photodestruction may be enhanced by up to a factor of a few in carbonate lakes by halide photoabsorption (\ref{sec:ap_timescales}). Saline systems have been disfavored as venues for prebiotic chemistry, on the argument that high salt concentrations generally inhibit lipid membrane formation (e.g., \citealt{Deamer2017}). However, some experimental work suggests that some amphiphile mixtures can form stable vesicles at oceanic salinity or higher, and indeed that some families of amphiphiles \textit{require} salinity for vesicle formation \citep{Namani2008, Maurer2016, Xu2017, Maurer2017}. We therefore argue that it is premature to dismiss saline waters as venues for prebiotic chemistry. Indeed, they may be favorable environments for the accumulation of molecules whose concentrations on early Earth are limited by $\leq220$ nm photolysis.

Waters rich in ferrous iron compounds such as ferrocyanide were low-UV (UV-shielded) environments. Such waters may have been favorable environments for UV-avoidant prebiotic chemistries. For example, \citet{Pearce2017} propose that meteoritically-delivered nucleobases in prebiotic ponds under the assumption that prebiotic pond-water was UV-opaque and could shield meteoritic nucleobases from photolysis. In a 1 m-deep lake with composition corresponding to the high-absorption endmember scenario for ferrocyanide lakes, the lifetimes of the nucleobases to direct photolysis would be enhanced by $\sim2$ orders of magnitude, making such waters potential candidates for the meteoritic delivery scenario (\ref{sec:ap_timescales}). By the same token, ferrous-rich lakes may be poor environments for UV-dependent prebiotic pathways. For example, the photoisomerization of diaminomaleonitrile (DAMN) to 5-aminoimidazole-4-carbonitrile (AICN), an intermediate for purine nucleobase synthesis, requires irradiation by $\lesssim300$ nm photons, which could be strongly attenuated in ferrocyanide-rich lakes \citep{Ferris1966, Cleaves2012, Boulanger2013, Yadav2020}. Similarly, the UV-driven deamination of cytidine to form uridine, which is driven by 200-300 nm photons could be inhibited in ferrocyanide-rich waters \citep{Powner2009, Todd2020}. Solvated electron production by UV photoirradiation of the ferrocyanide-sulfite system has been proposed to efficiently drive HCN homologation, but it is unknown whether ferrocyanide and/or sulfite is the photoactive agent, and if solely the latter, whether the chemistry can function with the dilute, optically thin ferrocyanide required for sulfite photolysis \citep{Xu2018, Ritson2018, Green2021}. We encourage further investigation into these questions in order to better understand the prebiotic potential of this chemistry. 



 In our discussion so far, we have focused on the implications of absorption of UV radiation by prebiotic solutes for direct photochemical processes like excitation or photolysis of biomolecules. However, the UV energy absorbed by prebiotic absorbers must be dissipated, and while it may be dissipated thermally, it may also be dissipated through the formation of high-energy species which can participate in further chemistry. For example, the halides, sulfite/sulfide, and ferrocyanide generate solvated electrons (e$^-_{aq}$) and oxidized radicals upon UV irradiation, which are reactive and may trigger further chemistry \citep{Jortner1964, Sauer2004a}. Such further chemistry may be destructive of biomolecules, meaning that even in waters with high concentrations of UV-absorbing compounds, UV light may indirectly suppress biomolecule accumulation. This condition must be checked by prebiotic chemistries invoking an aqueous UV "sunscreen". On the other hand, such further chemistry may also be productive. For example, \citet{liu2021prebiotic} report irradiation of SO$_3^{2-}$ and HCO$_3^-$ to lead to abiotic carbon fixation driven by e$^-_{aq}$ production from SO$_3^{2-}$, while the simultaneous irradiation of SO$_3^{2-}$ and ferrocyanide leads to HCN homologation \citep{Xu2018}. In these chemistries, the absorber effectively converts solar UV energy into chemical free energy in the form of e$^-_{aq}$. In sum, photochemical transformations may occur even in UV-shielded waters, triggered by photoabsorption of the very absorber which shields the UV; proposed prebiotic chemistry which invokes an aqueous-phase UV absorber must incorporate this possibility in verifying self-consistency.

\subsection{Caveats \& Limitations}
We have not considered FeOH$^+$ as a prebiotic sunscreen. FeOH$^+$ is a strong and broad UV absorber. Crucially, its absorption at the longer UV wavelengths where the Sun emits substantially more photons means that FeOH$^+$ is projected to have played a key role in Fe$^{2+}$ photooxidation and subsequent deposition \citep{Braterman1983, Anbar1992, Tabata2021}. Because of its ability to absorb longer-wavelength UV (300-450 nm), the total photoabsorption rate of FeOH$^+$ is comparable to that of Fe$^{2+}$ at circumneutral pH, despite much lower concentrations  \citep{Nie2017}. However, its absorption at longer wavelengths does not aid its potential as a sunscreen at shorter wavelengths (200-300 nm). At these wavelengths, the potential of FeOH$^+$ as a sunscreen is limited due to its low solubility at basic pH due to Fe(OH)$_2$ precipitation. This low solubility means that [FeOH$^+$] is low, implying large depths are required for it to significantly attenuate UV (i.e., bring the optical depth to unity). The solubility product of Fe(OH)$_2$ is $K_{sp}=4.9\times10^{-17}$, and $K_{eq}=3.02\times10^{-10}$ for the reaction $Fe^{2+} + H_2O \rightarrow FeOH^+ + H^+$ \citep{Braterman1983,CRC98}. For [Fe$^{2+}$]$=100~\mu$M, this implies [FeOH$^+$]$\leq2~\mu$M, restricting the relevance of FeOH$^+$ as an aqueous sunscreen for 200-300 nm radiation to deeper waters with $d>6$ m. 

We have considered in this work only a few of the vast diversity of natural waters. Other natural waters may have different absorption properties. For example, hot springs host high concentrations of HS$^-$, driven by continuous supply of H$_2$S from below \citep{Kaasalainen2011}. HS$^-$ is a strong and broad absorber, and like the ferrous iron compounds attenuates the longer-wavelength UV radiation which dominates solar UV output; some hot springs could therefore have been UV-opaque. Similarly, I$^-$ is a strong and broad UV absorber; in waters rich in iodine, e.g. mineral springs, I$^-$ may provide significant UV attenuation \citep{Fuge1986}.

We reiterate that the estimates of UV transmittance we present here are upper bounds. The longwave absorption of HCO$_3^-$ and CO$_3^-$ is unconstrained, and may be important in very alkaline waters. Prebiotic natural waters may have contained absorbers other than those we consider here, and if absorptive/abundant enough, these compounds may have served as sunscreens. Examples of such compounds include silicate or basaltic dust and meteoritically-delivered or atmospherically-derived organic compounds (B. Pearce, personal communication, 7/4/2019). Further work is required to determine the wavelength-dependent molar absorptivities of these alternate absorbers, and to estimate whether they could have been present in prebiotic waters at concentrations sufficient to significantly attenuate solar UV. Finally, the products of prebiotic chemistry may themselves attenuate UV \citep{Cleaves1998, Todd2021}. More detailed modelling of these products and their accumulation is required to rule on this possibility.

In this work, we have constructed the bracketing low- and high-absorptivity compositional endmembers for various prebiotic waters based on literature estimates. We have not attempted self-consistent geochemical modelling of these waters to estimate their composition, in part because the relevant kinetics have not been characterized under prebiotically-relevant conditions (e.g., \citealt{Ranjan2018, Ranjan2019}). Similarly, our work treats UV absorbers as static, and neglects further photochemical transformations triggered by their absorption of solar UV. Our work is therefore a first approximation, subject to revision by future models capable of modelling the self-consistent photochemistry and geochemistry of prebiotic natural waters. 

\section{Conclusions}
Prebiotic freshwaters may have been essentially transparent in the UV: UV-avoidant surficial prebiotic chemistries must invoke a UV "sunscreen" agent. Shortwave ($\leq220$) UV may have been attenuated in diverse prebiotic waters, with the most important absorbers being Br$^-$ in saline waters, and potentially sulfite in shallow lakes and nitrate in shallow closed-basin lakes. Better constraints on prebiotic NO$_X^-$ production rates and on prebiotic sulfite loss rates are required to rule on the latter possibility. Regardless, the impact of shortwave UV attenuation is modest, because most of the UV flux from the early Sun was delivered at longer, unshielded wavelengths, with the largest impact amongst the photoprocesses we considered being a modest enhancement in the lifetimes of the prebiotic intermediates 2-aminoxazole and 2-aminoimidazole to photolysis. Measurement of the wavelength-dependent rates of prebiotic processes are required to identify other processes that may be particularly promoted or inhibited in shortwave-shielded waters. The generally widespread availability of $\sim240-300$ nm radiation means that low-pressure mercury lamps remain suitable for initial studies of prebiotic chemistry, though more realistic irradiation is required to verify the plausibility of pathways discovered under mercury lamp irradiation. 

Some natural waters may have been largely opaque in the UV. In particular Fe$^{2+}$-derived compounds are effective broadband "sunscreens" if present at high concentrations. Ferrocyanide is an especially potent UV absorber, and ferrocyanide lake waters would have been low-UV environments which are candidate waters for UV-avoidant origin-of-life scenarios such as the meteoritic delivery hypothesis. On the other hand, such waters may be poor environments for UV-dependent reactions such as the photodeamination of cytidine to uridine. In addition to ferrocyanide, other ferrous iron compounds like FeSO$_4$ may serve as sunscreens in shallow lakes if present at proportionately higher concentrations to compensate for their lower molar absorptivities relative to ferrocyanide and to thermodynamically favor their complexation. Fe$^{2+}$ has been suggested to be widespread on early Earth, but the abundance and speciation of Fe$^{2+}$ compounds on early Earth remains uncertain; we highlight detailed modelling and/or geochemical constraints on ferrous iron concentrations and speciation in prebiotic natural (especially terrestrial) waters as a priority for UV-sensitive prebiotic chemistry. 

UV light can trigger photochemistry even if attenuated by an absorber, through photochemical transformations of that absorber. In particular, irradiation of a wide range of prebiotic absorbers, such as the halides, sulfite/sulfide, and ferrocyanide, is predicted to generate $e^-_{aq}$ and oxidized radicals. The implications of this production vary by prebiotic chemistry. To be plausible, prebiotic chemistries which invoke these absorbers must self-consistently account for the chemical effects of these by-products. The rates, timescales, and concentrations relevant for various prebiotic chemical reactions may also play a role in determining the overall self-consistency of different scenarios. We encourage further investigation in these areas to better understand the necessary environments and prebiotic plausibility of various origins-of-life chemistries.

\appendix

\acknowledgments
The authors are grateful to J. Cleaves, B. Pearce, J. Toner, J. Birkmann, Y. Beyad, J. Sutherland, J. Krissansen-Totton, S. Kadoya, L. Barge, and P. Rimmer for answers to questions and/or for discussions related to this paper. The authors further thank P. Rimmer and N. Greeen for comments on a draft of this paper, S. Kadoya for sharing the raw model outputs from \citet{kadoya2020}, and Y. Beyad for sharing the sulfite and bisulfite absorption spectra from \citet{Beyad2014}. The authors thank two anonymous referees whose feedback substantially improved this paper. The authors thank the Simons Collaboration on the Origin of Life and the Harvard Origins of Life Initiative for nurturing many fruitful conversations related to this paper. 

The raw data derived from the experiments, processed data reported in this paper, and scripts used to make the plots reported in this paper can be accessed on GitHub via \url{https://github.com/sukritranjan/dirty-water-1/}. 

\section*{Author Contributions}
S.R. conceived and lead study; G. L and C. L. K. performed measurements; C. L. K. synthesized data into molar decadic absorptivities and estimated errors; S. R. and A. H. performed modelling and literature data extraction for comparison; S. R., D. D. S and Z. R. T. explored prebiotic implications. 

\section*{Author Disclosures}
The authors are aware of no competing interests. 

\section*{Funding Statements}
This work was supported in part by grants from the Simons Foundation (SCOL Award\# 495062
 to S.R.; 3290360 to D. D. S.). Z.R.T. acknowledges support from the NASA Hubble Fellowship Program, award\# HST:HF2-51471.


%
%
\bibliography{dirtywater.bib}

\clearpage
\appendix
\section{Optical Properties of Pure Water}
\begin{figure}[h]
\noindent\includegraphics[width=0.8\textwidth]{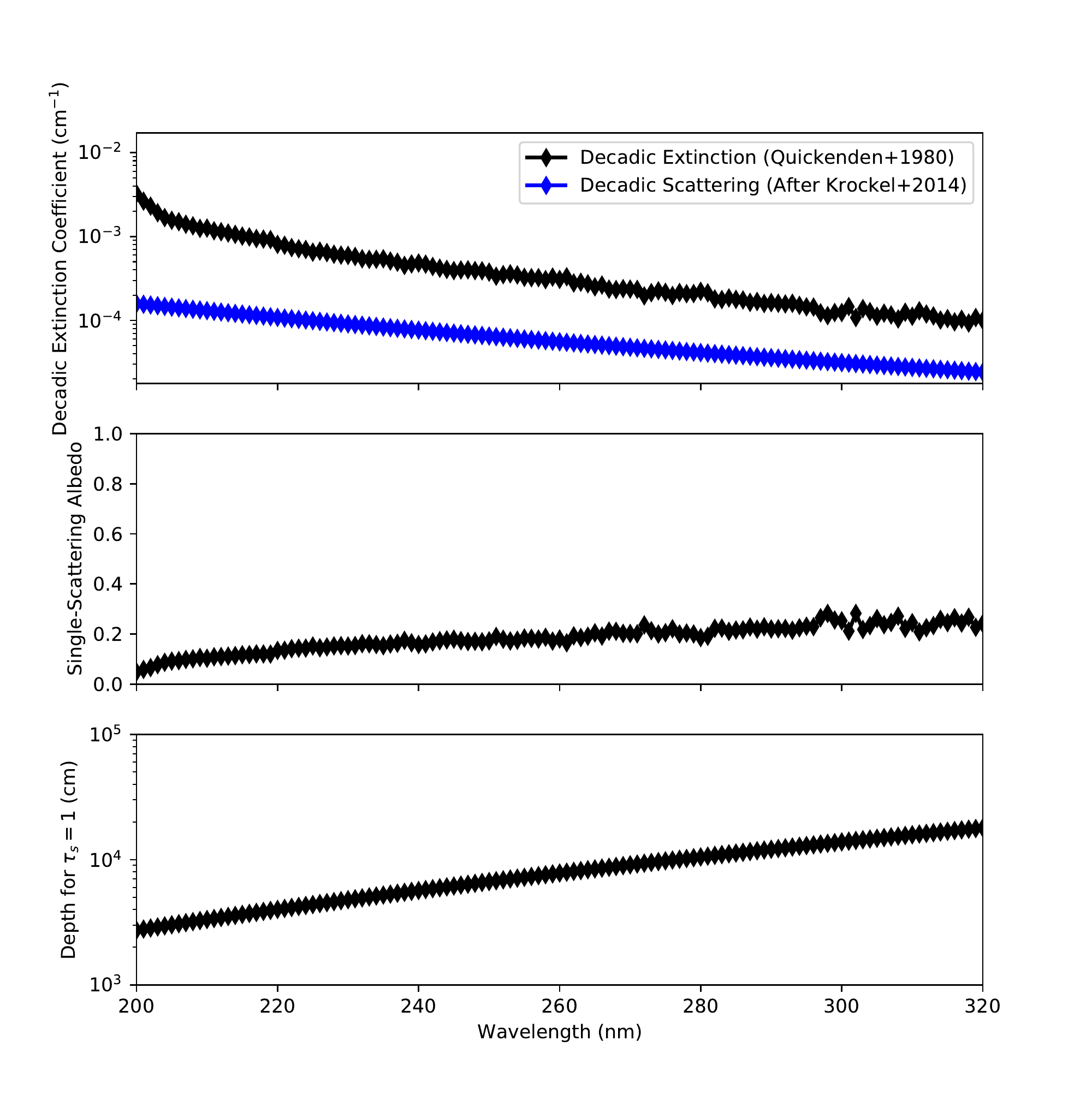}
\caption{Decadic extinction and scattering coefficients, single scattering albedo, and the depth at which the scattering optical depth $\geq 1$ as a function of wavelength for pure liquid water under standard conditions. The decadic extinction is taken from \citet{Quickenden1980}, which remains state-of-the-art in this wavelength range \citep{Krockel2014}. The Rayleigh scattering is calculated following \citet{Krockel2014}, with the modifications that the index of refraction, $n$, and the pressure derivative of $n$ at constant temperature $T$, $(\frac{\partial n}{\partial p})_T$, are taken to be constant at $n=1.39$ and $(\frac{\partial n}{\partial p})_T=0.095~\mu\text{m}^2~\text{kN}^{-1}$.} 
\label{fig:waterbeerlambert}
\end{figure}

\section{Molar Decadic Absorption Coefficients}
\subsection{Measured Molar Decadic Absorption Coefficients\label{sec:molabs_meas}}
\subsubsection{Samples} 
All salts were purchased from Sigma-Alrich (USA) at the highest available purity grade ($\geq$95\%) and used without further purification. All samples were dissolved in LC-MS grade freshwater (LiChrosolv, Millipore Sigma, USA). 

\subsubsection{Setup} 
The pH was measured by a pH electrode (Hach SensION + PH3, USA). The dissolved samples were kept in sealable spectrosil quartz cuvettes (9B-Q-10-GL14-C, Starna Cell’s, USA) with a sample depth of 10 mm. The concentrations of the samples were kept $\leq$0.1$\times$ the saturation concentration. Additionally the absorbance was kept between optical depth (OD) 0.005 and OD 1.1 throughout the spectral range of 200 nm to 360 nm by several dilution steps of the samples. All absorbance spectra were recorded in triplicates at 23$^\circ$C by a Shimadzu UV-1900 UV-VIS spectrophotometer relative to a blank water sample, and measurements were conducted at ambient conditions unless otherwise specified.

\subsubsection{Data evaluation}
After an initial solvent correction the triplicate spectra were averaged and cropped to an absorbance range between OD 0.005 and OD 1.1. The molar decadic absorption coefficients were calculated from the absorption spectra by division by the respective sample concentrations. Spectral ranges with systematic deviations, e. g. due to cuvette contamination, were excluded from the dataset. The systematic error of the spectrophotometric measurements was estimated conservatively to be about 15\%. Additionally, the following errors were included in the estimate: 6\% for the pipetting process, 4\% for the sample masses (14\% for NaS) and 5\% for sample impurities. In case of pH sensitive samples, an error of up to 4\% was included to account for the chemical equilibrium at the respective pH according to the Henderson-Hasselbalch equation. The statistical error for each sample was estimated from the averaging process. Table~\ref{tbl:sequence} gives the dilution sequence used to compile the molar absorptivity measurements. 

\begin{table}[h]
\caption{Measurement sequences used to determine molar decadic absorption coefficients. We report pH for the Fe(BF$_4$)$_2$ and K$_4$[Fe(CN)$_6$] measurement sequences to verify their speciation (Section~\ref{sec:concerns}). FeSO$_4$, FeCl$_2$, and Na$_2$Fe(CN)$_5$NO are not listed here because molar decadic absorption coefficients for those species were drawn from the literature. \label{tbl:sequence}}
\centering
\begin{tabular}{p{2.6cm}p{7.7cm}}
\hline
Sample  &  Measurement Sequence \\
\hline
NaBr  &  80 $\mu$M, 400 $\mu$M, 860 $\mu$M, 2 mM, 20 mM, 100 mM\\
KBr  & 70 $\mu$M, 350 $\mu$M, 1 mM, 20 mM, 100 mM  \\
NaCl  & 50 mM, 559 mM  \\
NaI  & 50 $\mu$M, 100 $\mu$M, 2 mM, 50 mM, 100 mM \\
KI  & 100 $\mu$M, 2 mM, 100 mM\\
Fe(BF$_4$)$_2$  & 1 mM (pH=3.4),  2 mM (pH=3.2), 20 mM (pH=2.6), 50 mM  (pH=2.3), 200 mM (pH=1.7) \\
K$_4$[Fe(CN)$_6$]  & 44 $\mu$M  (pH=6.5), 87 $\mu$M  (pH=6.4), 174 $\mu$M  (pH=6.4), 436 $\mu$M  (pH=6.0) \\
\\
\hline
\end{tabular}
\end{table}


\subsubsection{Experimental Concerns \& Mitigations\label{sec:concerns}}
We conducted our measurements at ambient conditions, i.e. in oxic air. Under oxic conditions ferrous iron slowly oxidizes. To mitigate this effect, each measurement sequences involving ferrous iron species were completed in $\leq1$ hour. We performed control experiments for the ferrous iron species under reduced oxygen conditions. We did not observe differences in the absorption spectra within the duration of our experiments, and in both cases found good agreement with the literature, suggesting that this procedure did not impact the accuracy of our approach. 

Special care was taken during the cuvette cleaning process to avoid contamination. In particular, the cuvettes were cleaned solely with water; we did not clean with acetone as we discovered in preliminary experiments that absorptivity due to residual acetone contaminated the measurements. 

Fe$^{2+}$ can complex with OH$^-$ in basic solution \citep{King1998}. We therefore tracked the pH of the Fe(BF$_4$)$_2$ solutions to ensure they remained at the acid pH at which Fe$^{2+}_{aq}$ is the dominant form. Conversely, in acidic solution Fe(CN)$_6^{4-}$ can protonate. We tracked the pH of the ferrocyanide solutions to ensure they well exceeded pH$=4.2$, corresponding to the pKa of the first protonation of Fe(CN)$_6^{4-}$ \citep{Shirom1969}. 

\subsection{Measured Spectra of Prebiotically-Relevant Compounds \& Comparison to Literature Data\label{sec:meas-data}}

In this section, we compare our measured spectra to those extracted from the literature. The literature data typically do not include uncertainty estimates. However, \citet{Birkmann2018} note that in previous studies, the uncertainty on their measurements of molar absorption coefficients was 5\%, as determined by reproduction. We assume this uncertainty estimate of 5\% to apply to the measurements reported in \citet{Birkmann2018} as well. 

\subsubsection{Bromide (Br$^-$)}
\begin{figure}[H]
\noindent\includegraphics[width=0.7\textwidth]{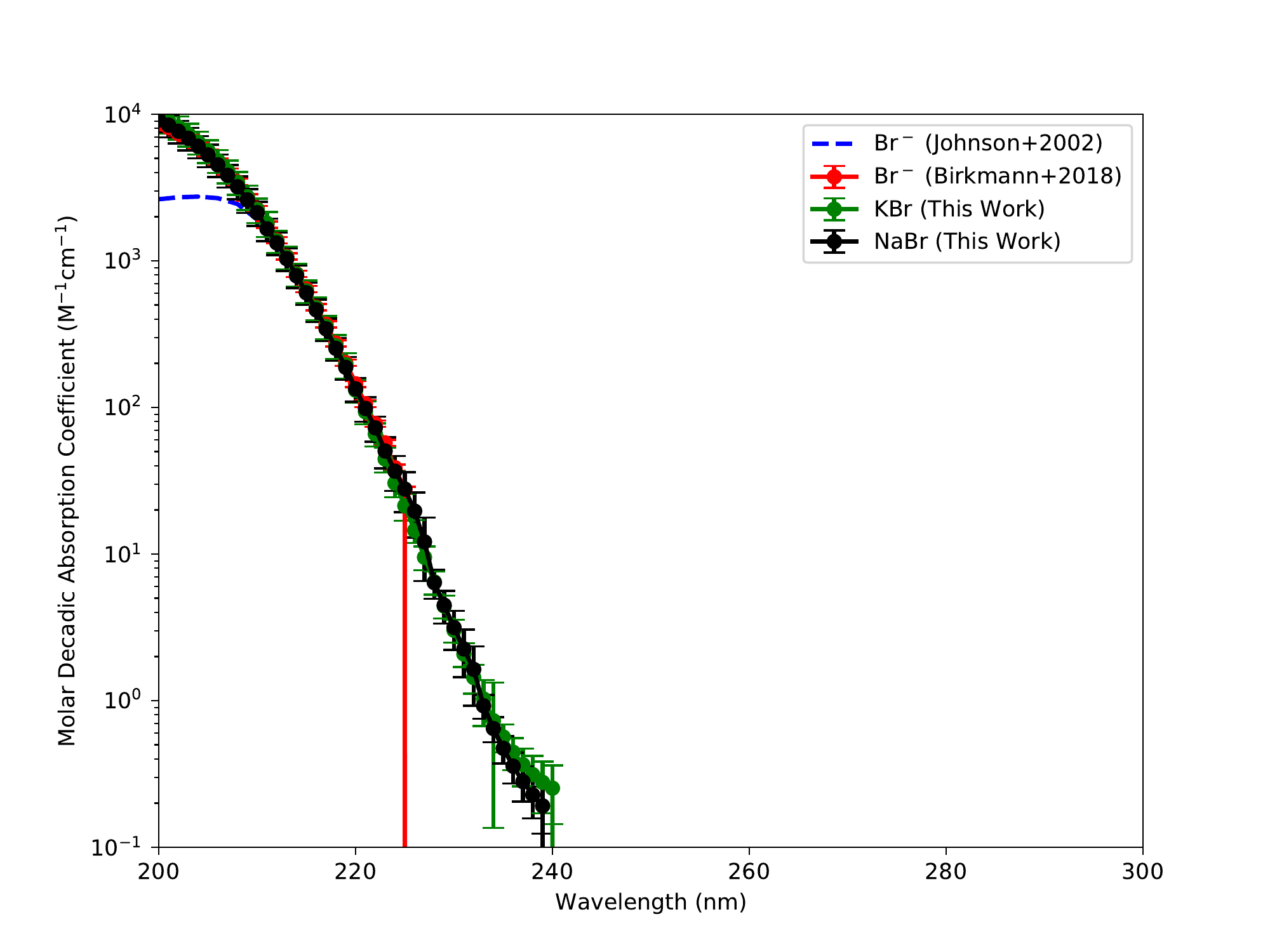}
\caption{Molar absorption coefficients of Br$^-$ derived from our measurements, compared to literature reports from \citet{Birkmann2018} and \citet{Johnson2002}. Our data extend spectral coverage of Br$^-$ absorption from 225 to 240 nm. We do not detect Br$^-$ absorption at $>$240 nm.}
\label{fig:Brm}
\end{figure}

Figure~\ref{fig:Brm} presents the molar absorption coefficients of NaBr and KBr derived from our measurements. The absorption of these species is identical to the precision of our measurements, confirming the findings of past work which indicate the anion dominates the absorption (\citealt{Birkmann2018} and sources therein). Our data agree with the measurements of \citet{Birkmann2018} from 200-225 nm, and extend the spectral coverage to 240 nm. We follow \citet{Birkmann2018} in disagreeing with \citet{Johnson2002} at short ($<204$ nm) wavelengths.

\subsubsection{Chloride (Cl$^-$)}

\begin{figure}[H]
\noindent\includegraphics[width=0.7\textwidth]{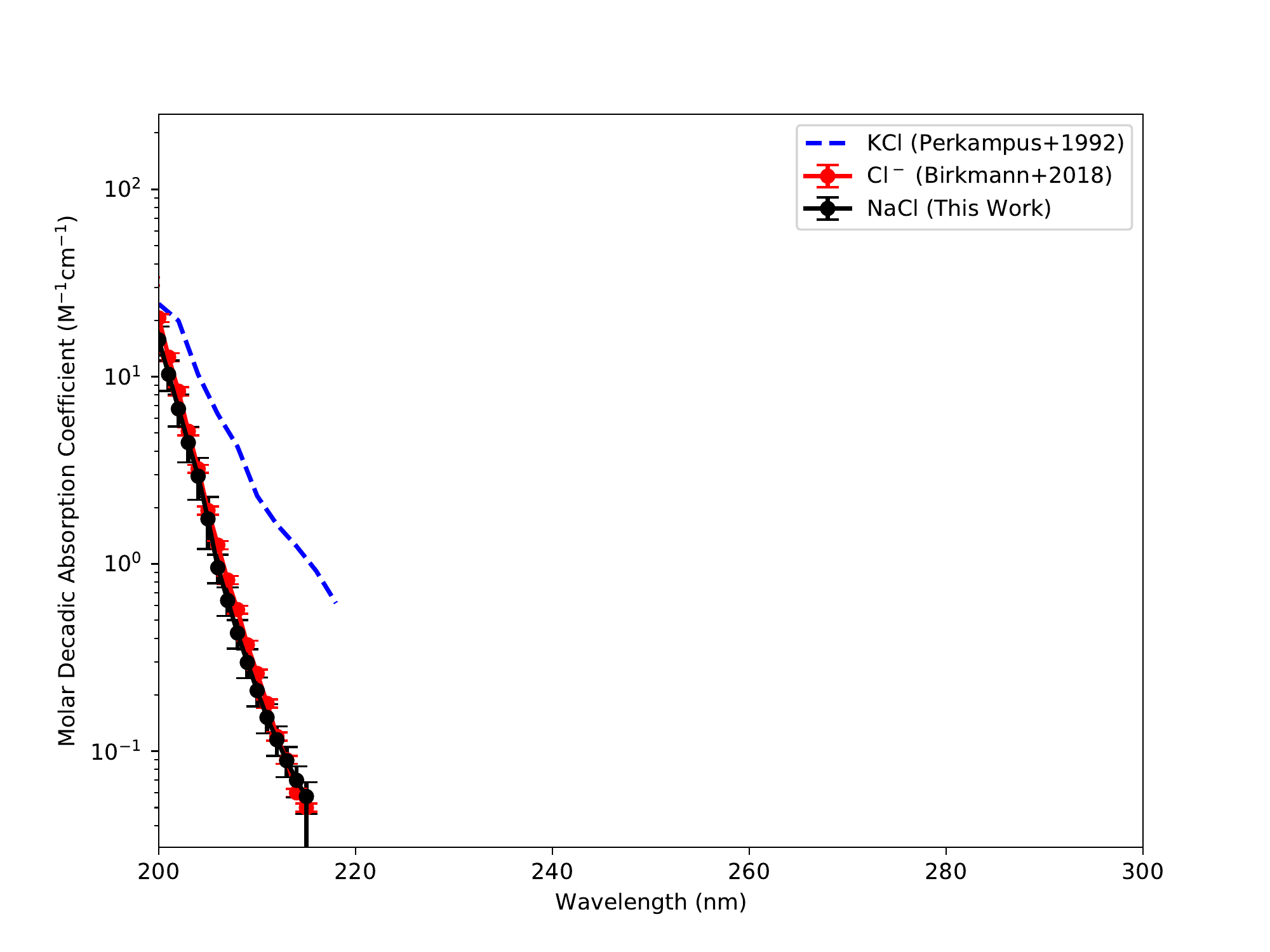}
\caption{Molar absorption coefficients of Cl$^-$ derived from our measurements, compared to literature reports from \citet{Birkmann2018} and \citet{Perkampus1992}. We do not detect Cl$^-$ absorption at $>$215 nm.}
\label{fig:Clm}
\end{figure}

Figure~\ref{fig:Clm} presents the molar absorption coefficients for NaCl derived from our measurements. We take the absorption of NaCl to be representative of Cl$^-$ generally, based on past work \citep{Birkmann2018}, and on our own measurements of KCl absorption. Our measurements agree with \citet{Birkmann2018} from 200-215 nm, and indicate that absorption of Cl$^-$ is much lower than reported by \citet{Perkampus1992}.

\subsubsection{Iodide (I$^-$)}
\begin{figure}[H]
\noindent\includegraphics[width=0.7\textwidth]{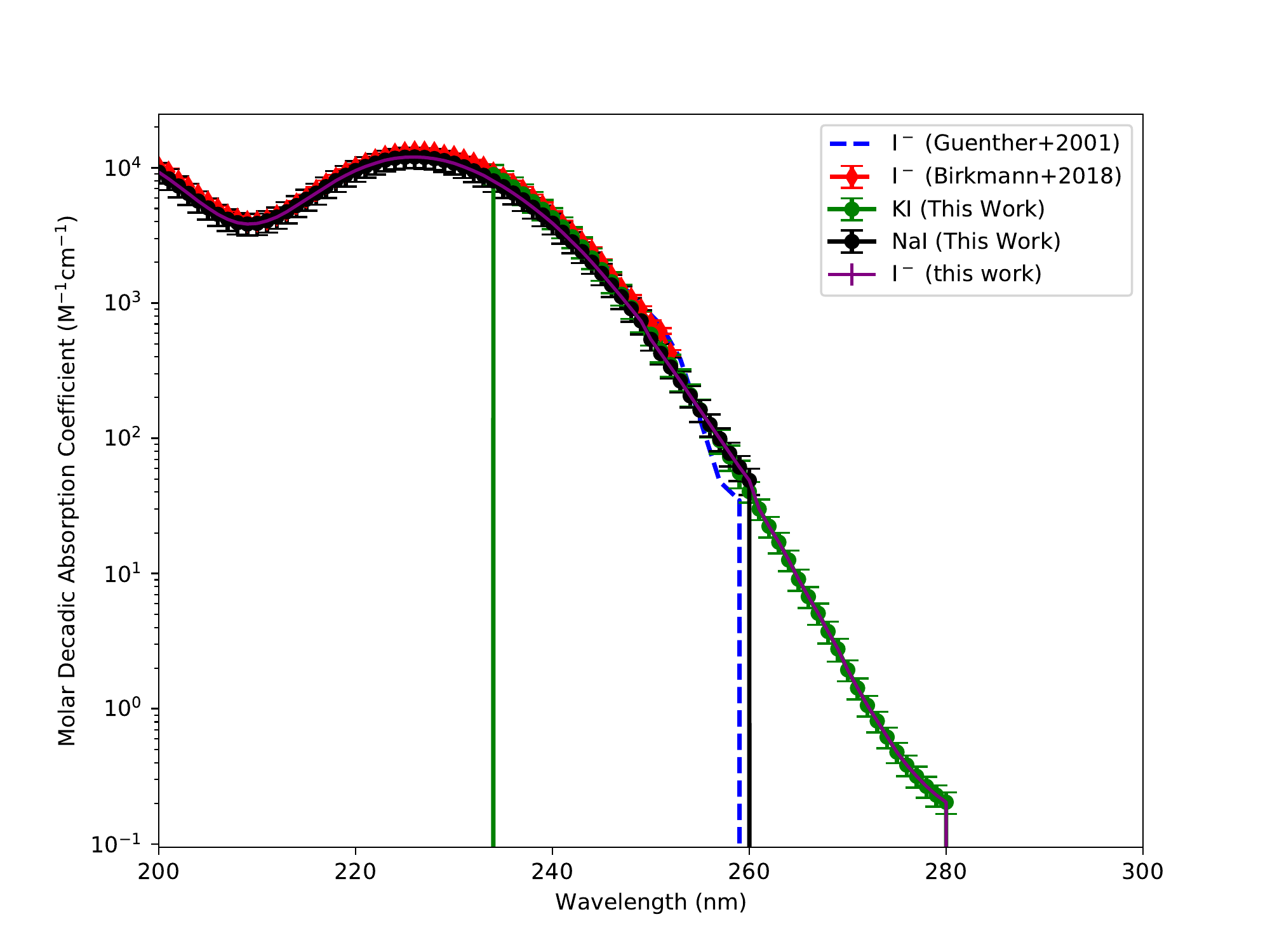}
\caption{Measured molar absorption coefficients of I$^-$, compared to literature reports from \citet{Birkmann2018} and \citet{Guenther2001}. Our data extend spectral coverage of I$^-$ absorption to 280 nm.We do not detect I$^-$ absorption at $>$280 nm.}
\label{fig:Im}
\end{figure}

Figure~\ref{fig:Im} presents the molar absorption coefficients of NaI and KI derived from our measurements. Where our data for these molecules overlap in spectral coverage, they are identical to within experimental uncertainty, consistent with past reports that the anion dominates UV absorption \citep{Birkmann2018}. We therefore combine the NaI and KI molar absorptivity measurements to synthesized a generalized I$^-$ absorption spectrum, which extend the longwave limit of spectral coverage of I$^-$ absorption from 252 nm to 280 nm. Our measurements agree with \citet{Birkmann2018} from 200-252 nm. Our measurements generally agree with \citet{Guenther2001}, but disagree slightly at their longest wavelengths of measurement.

\subsubsection{Fe$^{2+}$, as Fe(BF$_4$)$_2$\label{sec:fe2-in-isolation}}

We represent the absorption of Fe$^{2+}_{aq}$ by the absorption spectrum of Fe(BF$_4$)$_2$. Other workers (e.g., \citealt{Tabata2021}) instead take the absorption spectrum of Fe$^{2+}_{aq}$ from that reported by \citet{Heinrich1990}. \citet{Heinrich1990} derive their Fe$^{2+}$ by reducing iron metal (Fe$^0$) with HCl. This means that Cl$^-$ is present as a counter-ion in this solution, which influences the complexation and hence the UV spectrum of Fe$^{2+}$ . Most significantly, use of Cl$^-$ as a counter-ion leads to higher longwave UV absorption, and lower shortwave absorption \citet{Fontana2007}. Comparing the UV molar absorption coefficients of solutions of FeCl$_2$, FeSO$_4$, Fe(NH$_4$SO$_4$)$_2$, and Fe(BF$_4$)$_2$, \citet{Fontana2007} report solutions of Fe(BF$_4$)$_2$ to have the fewest bands attributable to complexes other than the Fe(H$_2$O)$_6^{2+}$ expected to form from Fe$^{2+}_{aq}$ in isolation. We therefore follow \citet{Fontana2007} in not using the absorption spectrum of FeCl$_2$ solution to represent the absorption spectrum of Fe$^{2+}_{aq}$ in isolation, and instead use Fe(BF$_4$)$_2$. In natural waters anions may in truth be present, which could enhance the absorptivity of the ferrous iron by complexation; the absorption of Fe$^{2+}$ therefore constitutes a lower bound on the UV absorption of ferrous compounds. 

\begin{figure}[H]
\noindent\includegraphics[width=0.7\textwidth]{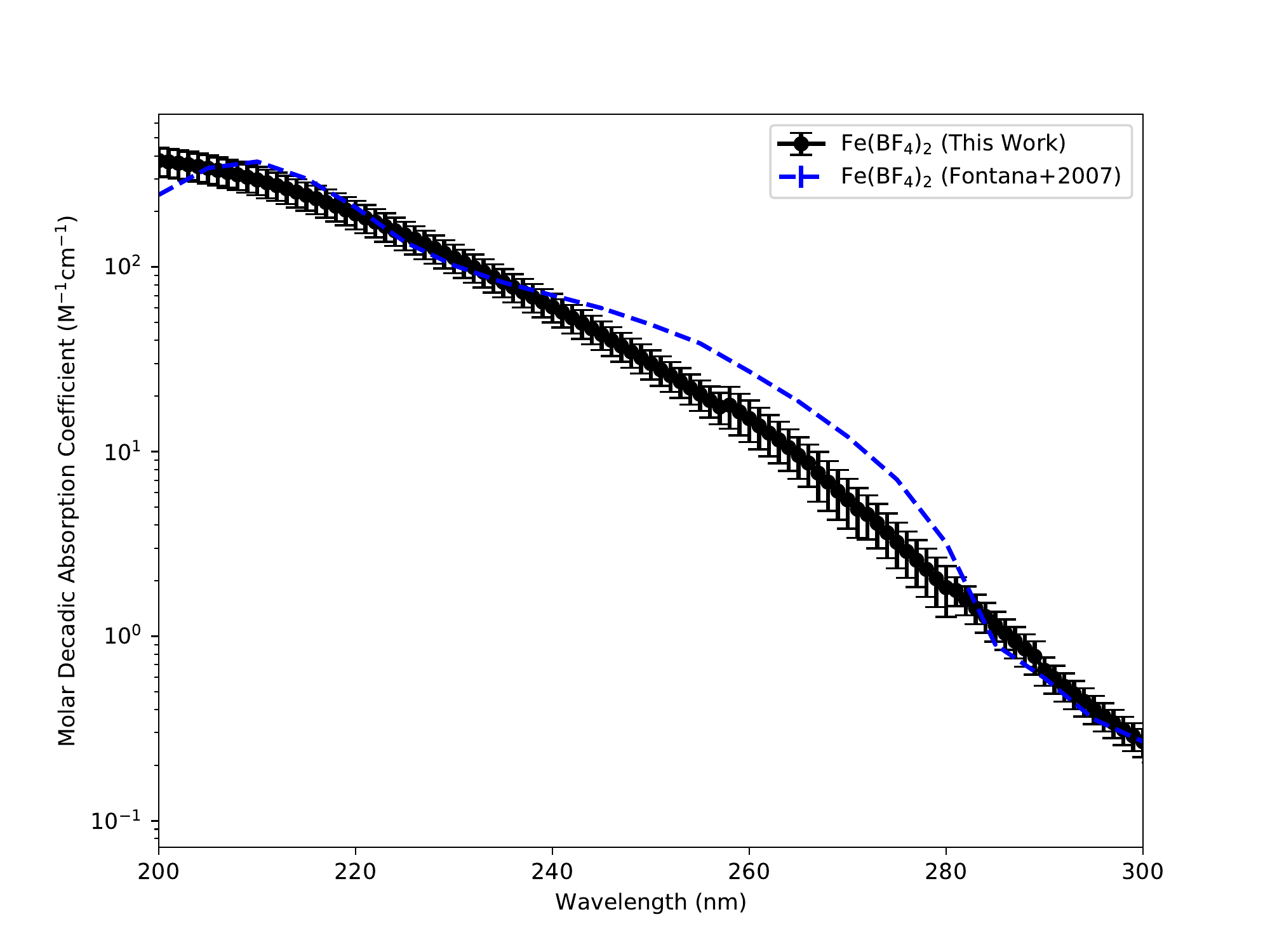}
\caption{Measured molar absorption coefficients of Fe(BF$_4$)$_2$, taken as representative of ferrous iron generally, compared to literature reports from \citet{Fontana2007}.}
\label{fig:FeBF42}
\end{figure}

Our measurement of the Fe(BF$_4$)$_2$ molar absorption coefficients in the UV is given in Figure~\ref{fig:FeBF42}. Our measurements generally agree with those extracted from \citet{Fontana2007}, but are slightly lower than theirs from 240-280 nm. Our measurements may therefore slightly underestimate UV absorption in this range. Our conclusions, which are at the order-of-magnitude level, are robust to this possibility. 

\subsubsection{Ferrocyanide}
\begin{figure}[H]
\noindent\includegraphics[width=0.7\textwidth]{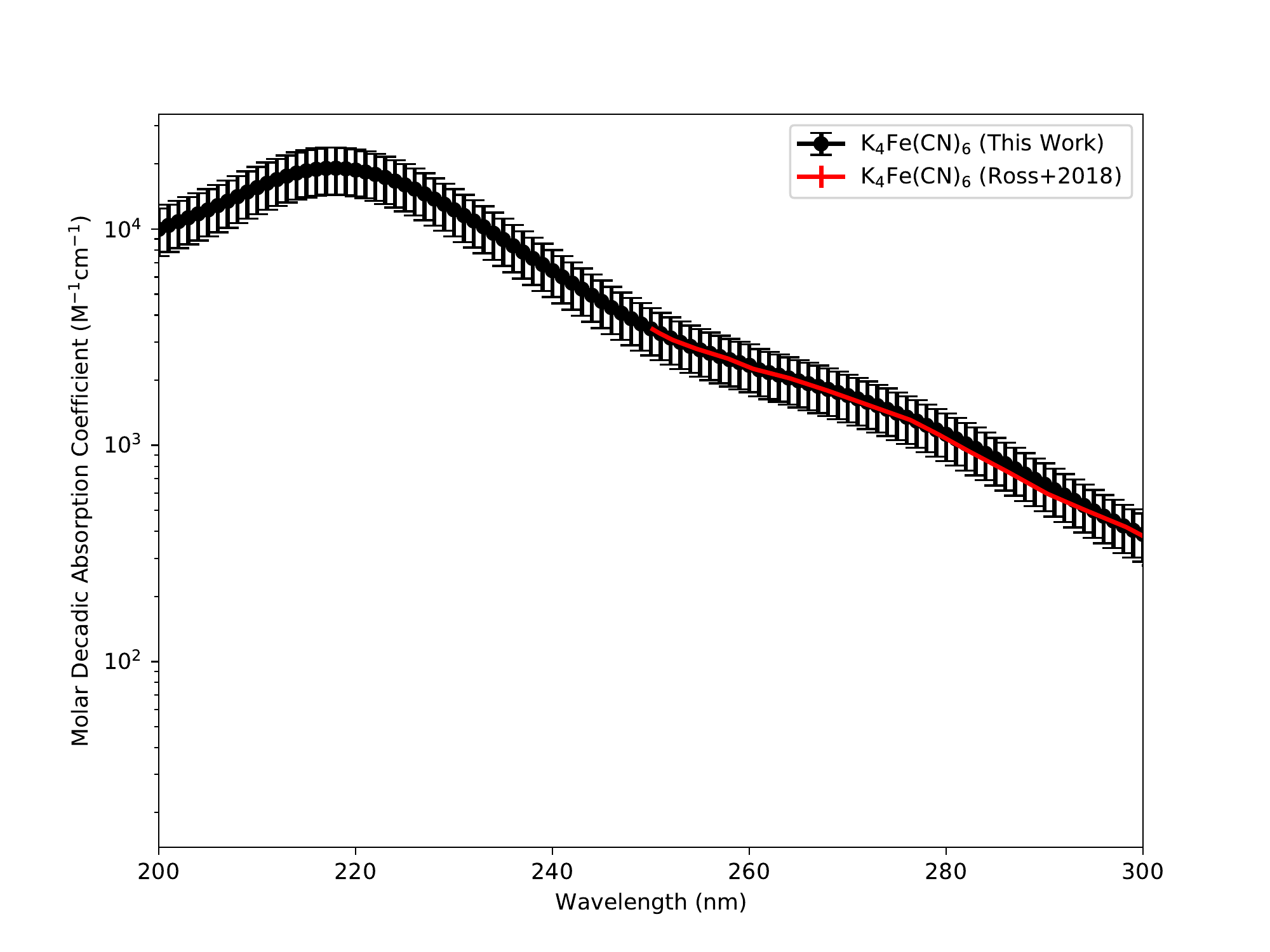}
\caption{Measured molar absorption coefficients of K$_4$Fe(CN)$_6$, taken as representative of ferrocyanide, compared to literature reports from \citet{Ross2018}. Our data extend spectral coverage of K$_4$Fe(CN)$_6$ absorption down to 200 nm.}
\label{fig:Ferrocyanide}
\end{figure}

 Figure~\ref{fig:Ferrocyanide} presents the molar absorption coefficents for K$_4$Fe(CN)$_6$ (ferrocyanide) derived from our measurements. The molar absorption coefficients we report agree with those extracted from \citet{Ross2018}, but we extend the spectral coverage blueward to 200 nm. 

\subsubsection{Nitrate}
\begin{figure}[H]
\noindent\includegraphics[width=0.7\textwidth]{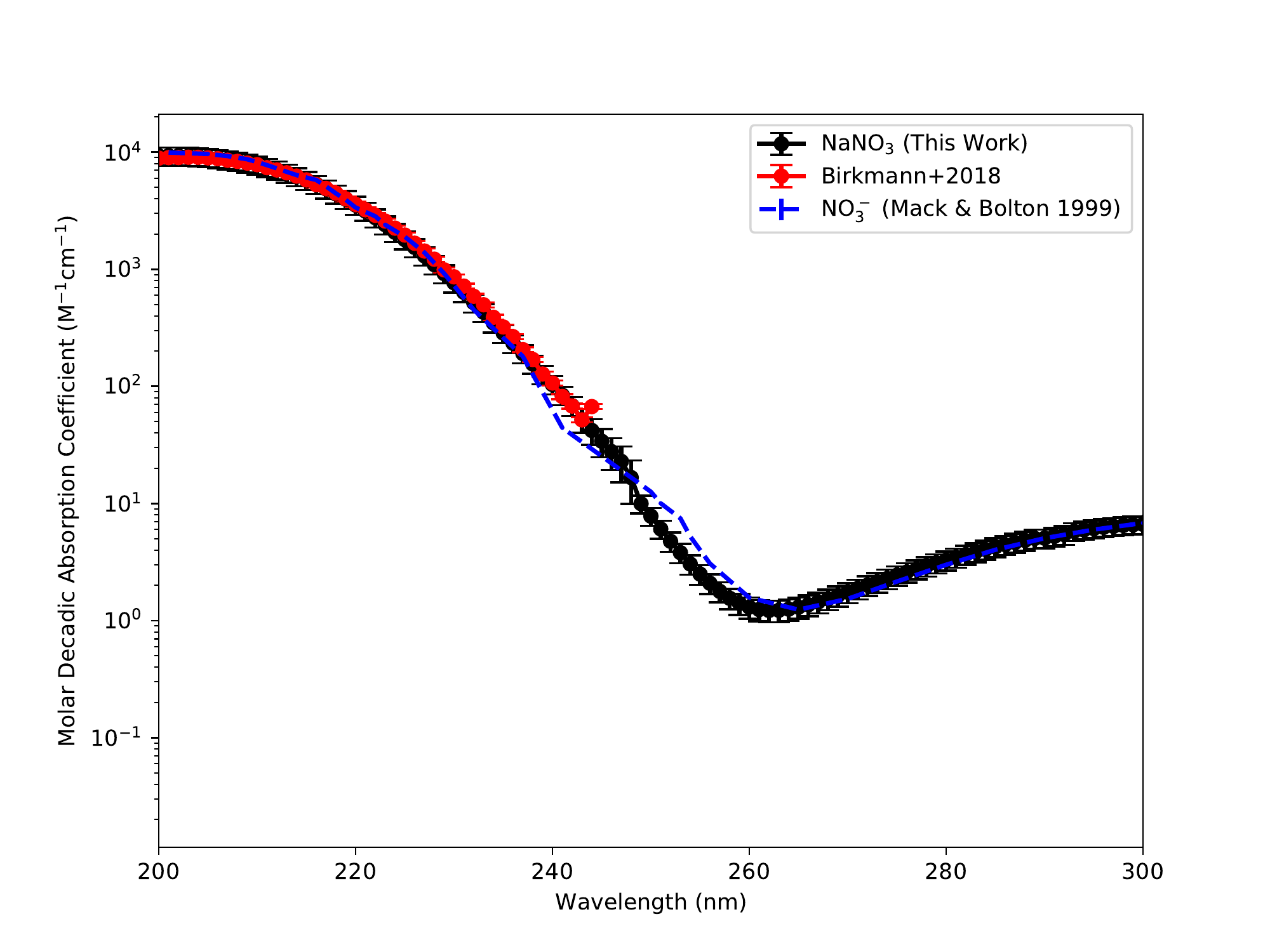}
\caption{Measured molar absorption coefficients of NaNO$_3$, taken as representative of the NO$_3^-$ ion, compared to the literature absorption spectrum of NO$_3^-$ of \citet{Mack1999} and \citet{Birkmann2018}. Our measurements agree well with the literature sources.}
\label{fig:nitrate}
\end{figure}

 Figure~\ref{fig:nitrate} presents the molar absorption coefficents for NaNO$_3$ (nitrate) derived from our measurements. The molar absorption coefficients we report are in reasonable agreement to those we extract from \citet{Mack1999} and \citet{Birkmann2018}.
 
\subsection{Literature-Derived Molar Decadic Absorption Coefficients \label{sec:molabs_lit}}

In this section, we show the molar decadic absorption coefficients we extracted from the literature. The literature data typically do not include uncertainty estimates. As in Section~\ref{sec:meas-data}, we assume an error of 5\% on the measurements reported by \citet{Birkmann2018}. We propagate this uncertainty under the assumption of uncorrelated, normally-distributed error, as with our own uncertainty estimates.

\subsubsection{Sulfite}
\begin{figure}[H]
\noindent\includegraphics[width=0.7\textwidth]{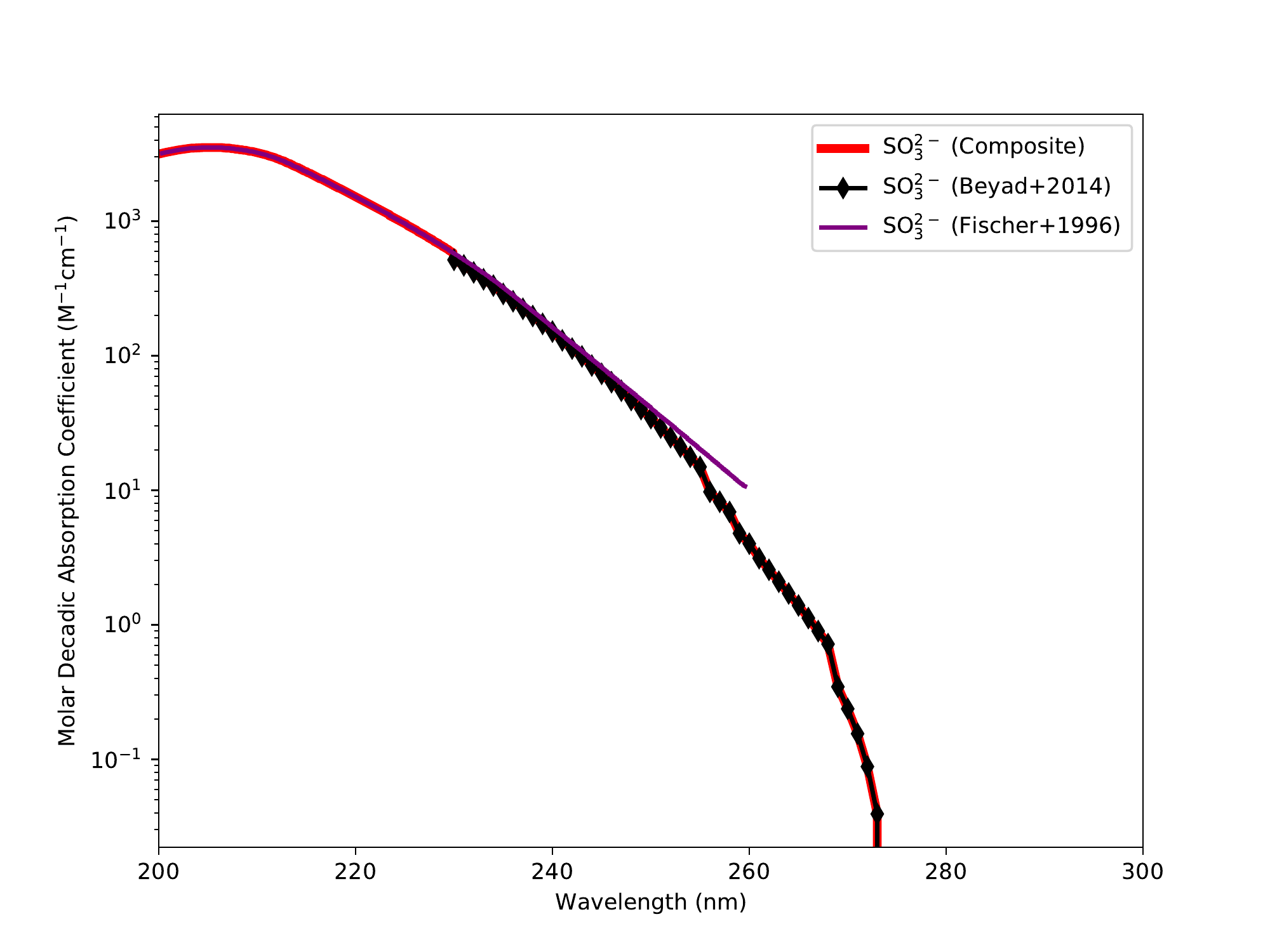}
\caption{Molar absorption coefficients of SO$_3^{2-}$, derived from \citet{Beyad2014} and \citet{Fischer1996}.}
\label{fig:so3mm}
\end{figure}

We combined the measurements of \citet{Beyad2014} and \citet{Fischer1996} to obtain an absorption spectrum for SO$_3^{2-}$. Specifically, we used the entire dataset of \citet{Beyad2014}, and concatenated the molar absorptivities of \citet{Fischer1996} at shorter wavelengths were data from \citet{Beyad2014} was unavailable (Figure~\ref{fig:so3mm}). 

\subsubsection{Bisulfite}
We combined the measurements of \citet{Beyad2014} and \citet{Fischer1996} to obtain an absorption spectrum for SO$_3^{2-}$. Specifically, we used the entire dataset of \citet{Beyad2014}, and concatenated the molar absorptivities of \citet{Fischer1996} at shorter wavelengths were data from \citet{Beyad2014} was unavailable. We used a log-linear extrapolation to connect the two datasets. (Figure~\ref{fig:so3mm}). 

\begin{figure}[H]
\noindent\includegraphics[width=0.7\textwidth]{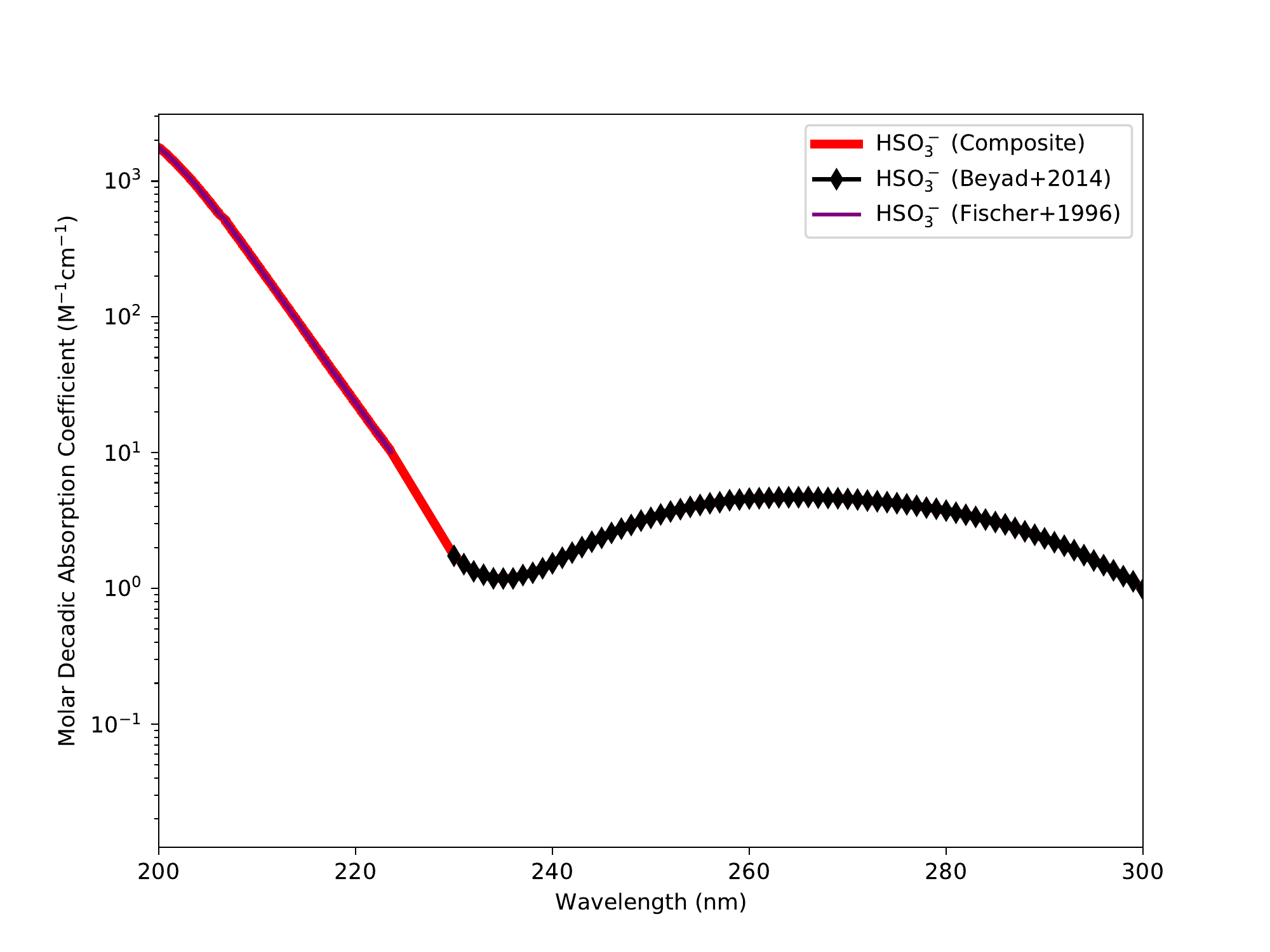}
\caption{Molar absorption coefficients of HSO$_3^{-}$, derived from \citet{Beyad2014} and \citet{Fischer1996}.}
\label{fig:hso3m}
\end{figure}

\subsection{Bisulfide}
\begin{figure}[H]
\noindent\includegraphics[width=0.7\textwidth]{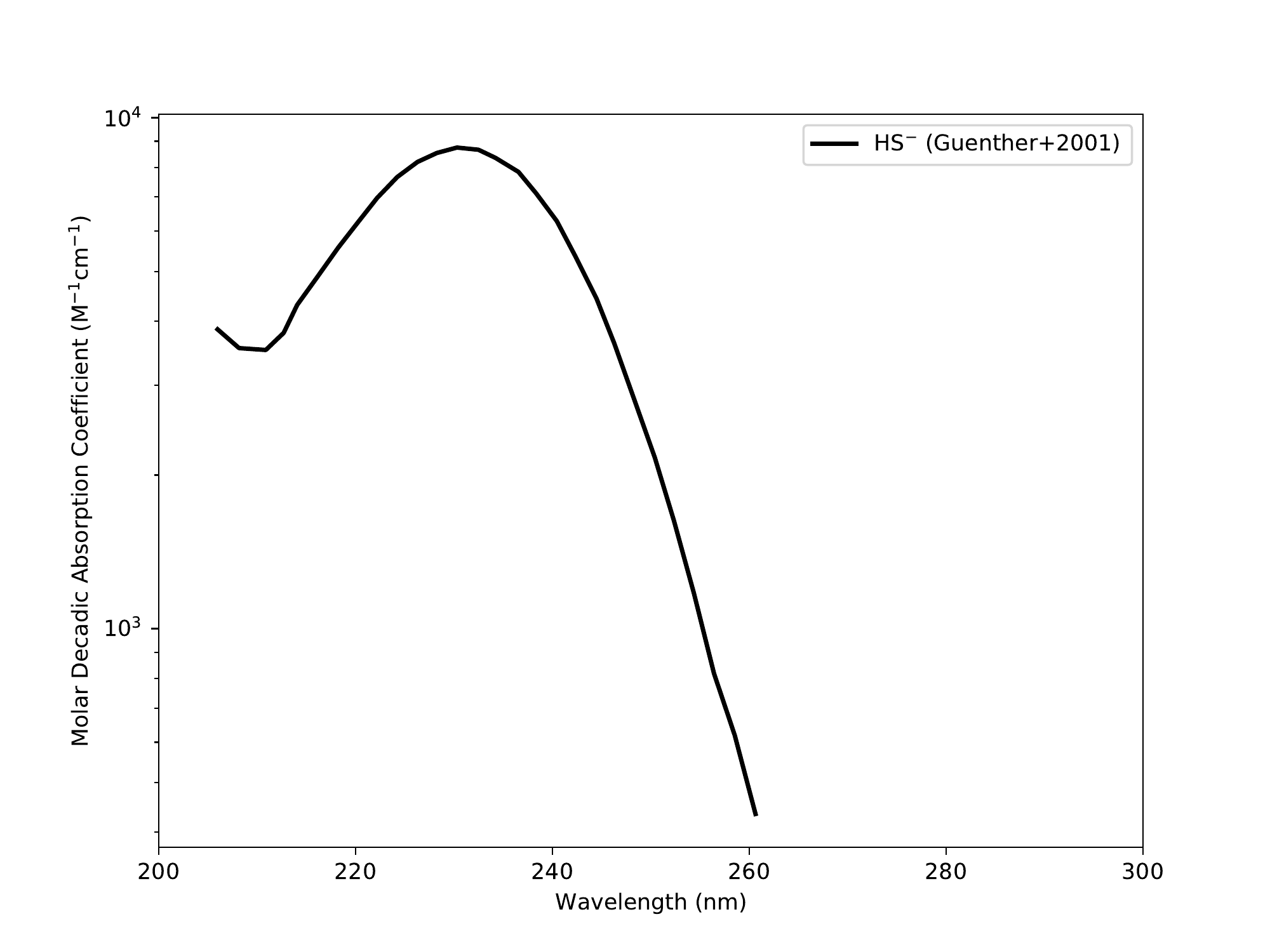}
\caption{Molar absorption coefficients of SO$_3^{2-}$, extracted from \citet{Guenther2001}.}
\label{fig:hsm}
\end{figure}

We derived molar absorption coefficients for HS$^-$ in the UV from the absorbance spectra of 50 $\mu$m HS$^-$ reported by \citet{Guenther2001} (Figure~\ref{fig:hsm}). These data terminate at $260$ nm, and our experimental procedure was inadequate to reliably constrain molar absorptions at longer wavelengths. We therefore caution that HS$^-$ may have additional absorption at longer wavelengths than we consider. 

\subsubsection{Carbonate and Bicarbonate}
\begin{figure}[H]
\noindent\includegraphics[width=0.7\textwidth]{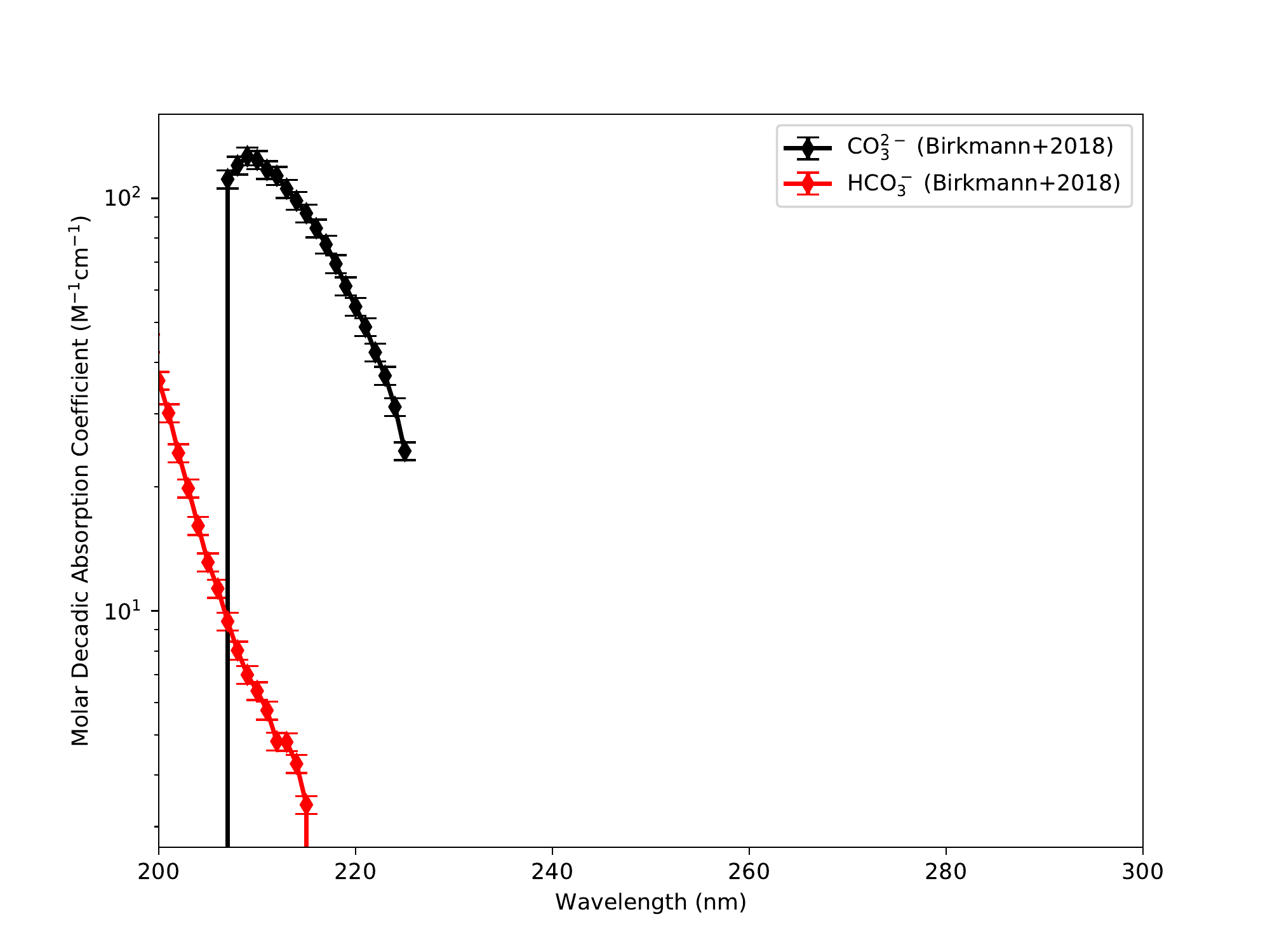}
\caption{Molar absorption coefficients of CO$_3^{2-}$ and HCO$_3^-$, from \citet{Birkmann2018}.}
\label{fig:DIC}
\end{figure}

We draw molar absorption coefficients for carbonate (CO$_3^-$) and bicarbonate (HCO$_3^-$) from \citet{Birkmann2018} (Figure~\ref{fig:DIC}). These measured spectra terminate abruptly at relatively short wavelengths. We are not aware of constraints on the absorption of bicarbonate and carbonate at longer wavelengths, and our efforts to reliably constrain absorption at these longer wavelengths were unsuccessful. We therefore caution that carbonate and bicarbonate may have additional absorption at longer wavelengths than we consider. 

\subsubsection{Other Fe$^{2+}$-Derived Compounds}
\begin{figure}[H]
\noindent\includegraphics[width=0.7\textwidth]{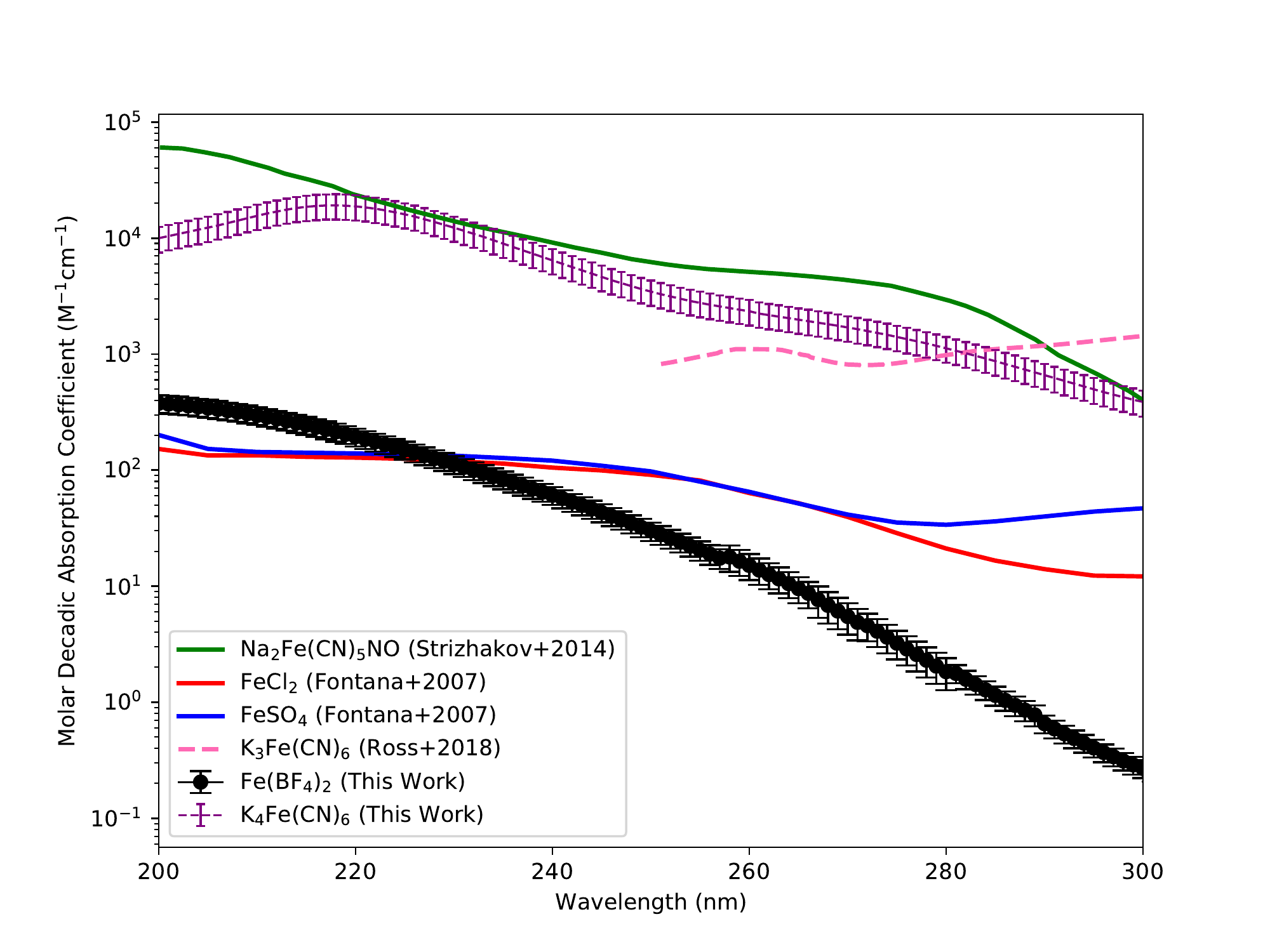}
\caption{Literature molar absorption coefficients of  Na$_2$Fe(CN)$_5$NO (sodium nitroprusside, SNP; \citealt{Strizhakov2014}), Fe(CN)$_6^{3-}$ (ferricyanide; \citealt{Ross2018}),  FeCl$_2$ \citep{Fontana2007}, and FeSO$_4$ \citep{Fontana2007}. Also shown are our derived molar absorption coefficients of K$_4$Fe(CN)$_6$ (ferrocyanide) and Fe(BF$_4$)$_2$, for context. The molar absorption of Fe$^{2+}$ varies dramatically depending on the anions with which it is complexed.}
\label{fig:SNP_etc}
\end{figure}

 FeCl$_2$ and FeSO$_4$ are potential complexes of Fe$^{2+}$ in natural waters, and nitroprusside and ferricyanide are potential photochemical derivates of ferrocyanide. We extract molar absorption coefficients of FeCl$_2$, FeSO$_4$, Na$_2$Fe(CN)$_5$NO (nitroprusside), and Fe(CN)$_6^{3-}$ (ferricyanide) from the literature \citep{Fontana2007, Strizhakov2014, Ross2018}. Iron and cyanide are extremely absorptive in the UV (Figure~\ref{fig:SNP_etc}).

\section{Estimating Photochemical Timescale Enhancement\label{sec:ap_timescales}}
Aqueous absorbers can decrease the rates of photochemical processes (e.g. photolysis, photoproduction), and hence increase their timescales, by attenuating the UV radiation which drives them. To gain an approximate sense of the magnitude of this effect for the prebiotic natural water endmember scenarios we study in this paper, we develop a simple formalism to crudely estimate the enhancement of timescales of photoprocesses in natural waters of nonzero absorbance. Specifically, we crudely estimate the enhancement in timescales of photochemical processes in well-mixed aqueous reservoirs by observing that the photochemical rate is linearly proportional to the UV flux \citep{Bolton2015}. Then, integrating the rate over the wavelengths of irradiation and the depth of the reservoir, and defining the timescale $T$ corresponding to a reaction rate $r$ to be $T=\frac{1}{r}$:

\begin{align}
    \frac{T}{T_0} &=\frac{\int_{\lambda_{1}}^{\lambda_{2}}  d\lambda~k(\lambda) \int_0^d dx}{\int_{\lambda_{1}}^{\lambda_{2}}  d\lambda~k(\lambda) \int_0^d dx~10^{-a(\lambda)x/\cos{\theta}}}\\
     & =\frac{d\int_{\lambda_{1}}^{\lambda_{2}}  d\lambda~k(\lambda) }{\int_{\lambda_{1}}^{\lambda_{2}}  d\lambda~k(\lambda) \int_0^d dx~10^{-a(\lambda)x/\cos{\theta}}}
\end{align}

Where $T_0$ is the photochemical timescale at the surface, $T$ is the photochemical timescale in the reservoir, $d$ is the depth of the reservoir, $a(\lambda) = \Sigma_i \epsilon_i(
\lambda) c_i$ is the decadic absorption coefficient of the reservoir as a function of wavelength, $\theta$ is the slant angle of the incident radiation, and $k(\lambda)$ is the wavelength-dependent photoreaction rate under early Earth surface conditions. We take $\theta=60^\circ$ ($\cos{\theta}=0.5$), which is consistent with both the angle of the diffuse stream in the two-stream radiative transfer formalism used to calculate the surface radiation field, and the assumed solar zenith angle in the atmospheric photochemical modelling on which our atmospheric transmission calculation is based \citep{Rugheimer2015, Ranjan2017a}.  $k(\lambda)$ may be (and generally is) a relative quantity, determined relative to a reference wavelength. $k(\lambda)$ have estimated for the photolysis of 2-aminooxazole, 2-aminoimidazole, and 2-aminothiazole, and for the photo-deamination of cytidine \citep{Todd2019, Todd2020}. 

We are not aware of a measurement of $k(\lambda)$ for the photolysis of the nucleobases; for this prebiotically important process, we crudely estimate the timescale following the simplifying assumption of \citet{Cleaves1998}, i.e. assuming that the photolysis rate is proportional to the irradiation at 260 nm, i.e. $ \frac{T}{T_0}=\frac{d}{\int_0^d dx~10^{-a(\lambda=260 nm)x}}$. We emphasize this approximation to be very crude; the peak absorbance of the biogenic nucleobases, while in the broad vicinity of 260 nm, often differs from 260 nm, and varies by nucleobase, pH, and structure of the incorporating molecule \citep{Voet1963}, and the wavelength-dependent quantum yield of photolysis of the nucleobases to our knowledge has not been strongly constrained in this wavelength region. This approximation should therefore be considered indicative only, and a call to detailed characterization of the wavelength-dependent nucleobase photolysis rates under early Earth conditions as conducted by \citet{Todd2019} for the 2-aminoazoles. 

Applying the above calculation, we estimate enhancements in the photolysis timescales of 2-aminooxazole and 2-aminoimidazole by a factor of a few ($2-5\times$) in closed-basin carbonate lakes (1 m-deep lake for the low-absorption endmember, 10 cm-deep lake for the high-absorption endmember). We estimate enhancements in the photochemical timescales of all 5 processes considered here by 2 orders of magnitude in the high-absorption endmember for the ferrocyanide lake scenario (1 m-deep lake). We emphasize the need for caution in extrapolating photochemical timescales to overall chemical timescales. Just because a process is photochemically inhibited does not mean that it does not proceed. For example, photoabsorption by Br$^-$ can generate Br radical \citep{Zafiriou1974}; if Br efficiently destroys 2-aminooxazole, then the overall lifetime of 2-aminooxazole is not necessarily enhanced in Br$^-$-rich waters despite inhibition of the direct photolysis pathway. Our calculations refer to specific photochemical processes; experiments are required to confirm whether the photochemical timescales are represenative of the overall timescales in realistic prebiotic mixtures.  

%
%
%
%
%

\end{document}